\documentclass[aps,twocolumn,superscriptaddress,showpacs,pra,twoside,amssymb,amsmath,latexsym]{revtex4}

\newcommand{\cP}{{\cal P}}
\newcommand{\bF}{{\bf F}}
\newcommand{\cE}{{\cal E}}
\newcommand{\rP}{{\rm P}}

\newcommand{\bX}{{\bf X}}

\newcommand{\sX}{\mathsf{X}}
\newcommand{\sZ}{\mathsf{Z}}
\newcommand{\sW}{\mathsf{W}}
\newcommand{\cD}{{\cal D}}

\def\vp{{\bf p}}

\def\vA{{\bf A}}
\def\vnu{\pmb{\nu}}
\def\vC{{\bf C}}
\def\vE{{\bf E}}
\def\vH{{\bf H}}

\def\complex{\mathbb{C}}

\def\Tr{\mathop{\rm Tr}\nolimits}
\def\Ker{\mathop{\rm Ker}\nolimits}
\def\im{\mathop{\rm Im}\nolimits}
\def\POS{\mathop{\rm POS}\nolimits}
\def\argmax{\mathop{\rm argmax}}
\def\argmin{\mathop{\rm argmin}}
\newcommand{\rE}{{\rm E}}
\newcommand{\dis}{{\rm dis}}

\newcommand{\cH}{{\cal H}}

\newtheorem{thm}{Theorem}
\newtheorem{prop}{Proposition}
\def\mix{{\rm mix}}

\def\choose#1#2{\genfrac{(}{)}{0pt}{}{#1}{#2}}

\begin{document}
\title{Upper bounds of eavesdropper's performances
 in finite-length code with decoy method}

\author{Masahito Hayashi}
\email{masahito@qci.jst.go.jp}
\address{ERATO-SORST Quantum Computation and Information Project, JST\\
5-28-3, Hongo, Bunkyo-ku, Tokyo, 113-0033, Japan}
\pacs{03.67.Dd,03.67.Hk,03.67.-a,05.30.Jp}
\begin{abstract}
Security formulas of quantum key distribution (QKD) with imperfect resources are obtained for finite-length code when the decoy method is applied. This analysis is useful for guaranteeing the security of implemented QKD systems. Our formulas take into account the effect of the vacuum state and dark counts in the detector. We compare the asymptotic key generation rate in presence of dark counts with that without.
\end{abstract}
\maketitle

\section{Introduction}
The BB84 protocol proposed by Bennett and Brassard\cite{bene} in 1984
attracts attention as an alternative to modern cryptography based on
complexity theory. Many efforts are devoted to searching for
implementations of quantum communication channels for this purpose.
The security of the original BB84 protocol can be trivially proved
only when the quantum communication channel is noiseless. Since 
there is noise in any implemented quantum channel, it is needed to prove the
security with the noisy channel, which has been proved by
Mayers\cite{mayer1}. After his proof, many different proofs were
reported. However, any implemented quantum channel, besides loss, also
suffers imperfections in generating a single photon. That is, the sent
pulse is given as a mixture of the vacuum state, the single-photon
state, and the multi-photon state, and it is impossible for the sender
(Alice) and the receiver (Bob) to identify the number of photon. In
order to guarantee the security in such a case, the decoy method has
been proposed\cite{hwang,LMC,XBZL,wang}, in which different kinds of
pulses are transmitted. However, these preceding researches did not
provide security with the finite-length code, which is a basic
requirement in practical settings. That is, there is no established
method to evaluate quantitively the security of an implementable
quantum key distribution (QKD) system.

On the other hand, modern cryptographic methods are required to
evaluate its security quantitively. Hence, for the practical use of
QKD, it is needed a theoretical analysis in order to present
quantitive criteria for security and to establish the method to
guarantee this criteria for the implemented QKD system. If nothing in
this direction is done, QKD systems cannot be developed for practical
use.

In a usual QKD protocol, the final key is generated via classical
error correction and privacy amplification after the initial key (raw
key) is generated by the quantum communication. In the classical error
correction part, it is sufficient to choose our classical error
correction code based on the detected error rate. Privacy
amplification, on the other hands, sacrifices several keys in order to
guarantee the security against the eavesdropper. The upper bound of
eavesdropper(Eve)'s information for the final key is closely related
to the amount of sacrifice bits.

Since Eve's information for the final key is the measure of the
possibility of eavesdropping, its quantitive evaluation is required.
In order to decrease Eve's information sufficiently, we need a
sufficient amount of sacrifice bits, which is given by the product
between the length of our code and the rate of sacrifice bits. A
larger size of our code requires larger complexity of the privacy
amplification, and a larger rate of sacrifice bits decreases the
generation rate of the final key. Hence, it is required to derive the
formula to calculate the upper bound of Eve's information for the
final key for a given length of the code and a given rate of sacrifice
bits, under the realizable quantum communication channel.

Our problems can be divided into three categories: The first is the
evaluation of Eve's information for the given length of our code and
the given rate of sacrifice bits. Since any implemented QKD system has
a finite-length code, any asymptotic security theory cannot guarantee
the security of an implemented QKD system. The second is the security
analysis for imperfect resource (e.g., phase-randomized coherent
light) that consists of mixtures of the vacuum state, single-photon
state, and multi-photon state. Many practical QKD systems are equipped
not with single-photon but with weak phase-randomized coherent
signals. These systems require a security analysis with an imperfect
resource. Further, even if a QKD system is approximately equipped with
single-photon signals, it nonetheless needs a security analysis for an
imperfect resource because only a perfect single-photon resource
allows the security analysis for the single-photon case. The third is
the identification of the relative ratio among the vacuum state, the
single-photon state, and the multi-photon state in the detected
pulses. Many implemented quantum communication channels are so lossy
that Alice and Bob cannot identify this ratio 
in the detected pulses 
even though they know this ratio in the transmitted pulses.
Thus, they need a method to estimate this ratio. Each of these three
problems has been solved only separately, however, an implemented QKD
system requires a unified solution for these three problems, which
cannot be obtained by a simple combination of separate solutions.

Concerning the first problem, many papers treated only the asymptotic
key generation (AKG) rate.  Only the papers, Mayers\cite{mayer1},
Inamori-L\"{u}tkenhaus-Mayers(ILM)\cite{ILM},
S.Watanabe-R.Matsumoto-Uyematsu(WMU)\cite{WMU},
Renner-Gisin-Kraus(RGK)\cite{RGK}, and Hayashi\cite{hayashi} discussed
the security of the finite-length code with a low complexity protocol.
In particular, only ILM\cite{ILM} takes into account the second
problem among them, and the other papers treat only the single photon
case. Extending the method of Mayers\cite{mayer1}, ILM \cite{ILM}
provided an evaluation of the security with imperfect resources for
the finite-length code. Unfortunately, their formula for the security
evaluation is so complicated that a simpler security bound is needed.
They also obtained the AKG rate with imperfect resources. Extending
the method of Shor-Preskill\cite{SP},
Gottesman-Lo-L\"{u}tkenhaus-Preskill (GLLP)\cite{GLLP} also obtained
this rate. In order to solve the third problem, Hwang\cite{hwang}
developed the decoy method, in which we estimate the ratio by changing
the intensity of the transmitted phase-randomized coherent light
randomly. After this breakthrough, applying the asymptotic formula by
GLLP\cite{GLLP}, Wang\cite{wang} and Lo et al.\cite{LMC,XBZL} analyzed
this method deeply, but did not treat the security of the
finite-length code. Hence, there is not enough results to treat the
security with the decoy method for the finite-length code.

Further, there is a possibility for an improvement of the AKG rate by
ILM \cite{ILM} and GLLP\cite{GLLP}. Taking into account the effect of
the vacuum state, Lo\cite{Lo} conjectured an improved AKG rate.
Considering the effect of the dark counts in the detector,
Boileau-Batuwantudawe-Laflamme (BBL)\cite{BBL} conjectured a further
improvement of the AKG rate conjectured by Lo\cite{Lo}. They pointed
out that the AKG rate with the forward error correction is different
from that with the reverse error correction.

In this paper, 
in order to evaluate eavesdropper's performances,
we focus on the average of Eve's information,
the average of the maximum of trace norm between Eve's states
corresponding to different final keys,
and the probability that Eve can correctly detect the final key.
We derive useful upper bounds of these quantities
for the protocol given in section \ref{s2}
in the finite-length code by use of mixing different imperfect resources.
Based on this bound, we obtain an AKG rate. 
In particular, due to the consideration of the effect of
the dark counts, our bound improves that by ILM \cite{ILM}, and it
yields the AKG rate that coincides with the that conjectured by
BBL\cite{BBL}. We should mention here that our description for quantum
communication channel is given as a TP-CP map on the two-mode bosonic
system. Since our results can be applied to the general imperfect
sources, it provides security with an approximate single-photon
source. However, further statistical analysis is required for the
numerical bound of Eve's information for implemented QKD system with
the finite-length code. Such an analysis is presented in another paper
\cite{H3}. Also, the analysis of the AKG rate in the case of
phase-randomized coherent light will be presented in another paper
\cite{H2}.

The paper is organized as follows. In section \ref{s2}, as a
modification of BB84 protocol, we present our protocol, in which we
clarify the measuring data deciding the size of sacrifice bits in the
privacy amplification. In section \ref{s3}, we derive upper bounds
of the averages of Eve's information about the final key and of the
trace norm of the maximum between Eve's states corresponding to
different final keys under the protocol given in section \ref{s2}. In
section \ref{s4}, we characterize the AKG rate based on our bounds,
and apply it to the case of mixture of the vacuum and the single-photon 
and the case of approximate single-photon. In Section \ref{s5}, the quantum
communication channel is treated as a general TP-CP map on the
two-mode bosonic system. It is proved that such a general case can be
reduced to the case given in section \ref{s3}.

\section{Modified BB84 protocol with decoy state}\label{s2}
We consider BB84 protocol based on 
$+$ basis, $|\uparrow\rangle$, $|\downarrow\rangle$ 
and $\times$ basis,
$|+\rangle:= \frac{1}{\sqrt{2}}(|\uparrow\rangle+|\downarrow\rangle)$, 
$|-\rangle:= \frac{1}{\sqrt{2}}(|\uparrow\rangle-|\downarrow\rangle)$.
If we realize this protocol by using photon (or bosonic particle),
we have to generate single-photon in the two-mode system
and transmit it without no loss.
However, it is impossible to implement this protocol perfectly,
any realized quantum communication system can send only an imperfect photon
(or approximately single-photon).
Hence, we have to treat bosonic system more carefully.
Let us  give its mathematical description.
Two-mode bosonic system is described by
$\cH:=\oplus_{n=0}^{\infty} \cH_n$,
where
the $n$-photon system
$\cH_n $ is the Hilbert space spanned by 
$|0,n\rangle, |1,n-1\rangle,\ldots, |n-1,1\rangle$, and $|n,0\rangle$.
For example, $|j,n-j\rangle$ is the state consisting of 
$j$ photons with the state $|\uparrow\rangle$ 
and $n-j$ photons with the state $|\downarrow\rangle$.
Also the vector $\sum_{j=0}^{n}\sqrt{
\choose{n}{j}(\frac{1}{2})^n}
|j,n-j\rangle$
($\sum_{j=0}^{n}\sqrt{
\choose{n}{j}(\frac{1}{2})^n}
(-1)^{n-j}
|j,n-j\rangle$)
corresponds to the state 
of $n$ photons with the state 
$|+\rangle$ ($|-\rangle$).
That is, 
the system $\cH_n $ is equivalent with
the $n$-th symmetric subspace of two dimensional system.
We also denote 
the state of $j$ photons with the state $|+\rangle$ 
and $n-j$ photons with the state 
$|-\rangle$ by $|j,n-j,\times\rangle$.

When we would generate the state 
$|\uparrow\rangle$ in the two-dimensional system
with the coherent pulse,
the generated state is 
described by the state $\sum_{n=0}^{\infty}e^{-\frac{|\alpha|^2}{2}}
\frac{\alpha^n}{\sqrt{n!}}|n,0\rangle$
in the two-mode bosonic system.
However, 
if we implement our system so that
each phase factor $\theta$ of
the complex amplitude 
$\alpha= \sqrt{\mu}e^{i\theta}$ 
is completely random,
our state can be regarded as
the mixed state 
$e^{-\mu}\sum_{n=0}^\infty \frac{\mu^n}{n !}
|n,0 \rangle \langle n,0|$,
which depends only on the intensity $\mu$.
In the following, 
we consider a more general case, in which 
the pulse sent by the sender (Alice) are given by
$\rho_{0,+}^{\nu}:=
\sum_{n=0}^\infty \nu(n)|n,0 \rangle \langle n,0|$,
$\rho_{1,+}^{\nu}:=
\sum_{n=0}^\infty \nu(n)|0,n \rangle \langle 0,0|$,
$\rho_{0,\times}^{\nu}:=
\sum_{n=0}^\infty \nu(n)|n,0 ,\times\rangle \langle n,0,\times|$,
$\rho_{1,\times}^{\nu}:=
\sum_{n=0}^\infty \nu(n)|0,n ,\times\rangle \langle 0,n,\times|$,
where $\nu$ is an arbitrary distribution.

Since our communication channel is lossy,
the receiver (Bob) cannot necessarily detect all of the sent pulses.
If the breakdown of the detected pulses 
(the ratio among 
the vacuum state, the single-photon state, 
$n$-photon state, and so on)
is known,
we can guarantee the security of BB84 protocol
based on the discussion on subsection \ref{2-16-1}.
However, since the usual quantum communication channel is lossy,
there is a possibility that Eve can 
control the loss depending on the number of the photons.
Hence, it is impossible to identify the the loss of each 
number of the photons if Alice sends the pulse by using 
one distribution $\nu$.
One solution is the decoy method\cite{hwang,wang,LMC},
in which Alice randomly chooses the distribution $\nu$
and estimates the loss and the error probabilities
of each number state.
It is effective to choose the vacuum pulse 
$|0 \rangle \langle 0|$.

In the following, we describe our protocol.
First, we fix the following;
the size $N$ of our code,
the maximum number $\overline{N}$ and
the minimum number $\underline{N}$ 
of final key size,
the number $N'$ of sent pulses, 
the $k$ distributions 
$\nu_1,\ldots, \nu_k$ of the generated number of photons,
and 
the distribution $\nu_{i_0}$,
whose pulse generates the raw keys.
Since the vacuum pulse and two bases are available,
Alice sends $2k+1$ kinds of pulses,
where 
the $0$-th kind of pulse means the vacuum pulse,
the $i$-th kind of pulse means the pulse with
the $\times$ basis generated by the distribution $\nu_i$,
and 
the $i+k$-th kind of pulse means the pulse with
the $+$ basis generated by the distribution $\nu_i$
for $i=1,\ldots, k$.
For this purpose, they fix the probabilities
$\overline{p}_0,\ldots, \overline{p}_{2k}$,
and Alice generates 
$i$-th kind of pulse with the probability $\overline{p}_i$
for $i=0,\ldots, 2k$.
The probabilities $\overline{p}_{i_0}$
and $\overline{p}_{k+i_0}$ should be larger
because these generate the pulses producing the raw keys.
In this paper, we use the bold style for describing the vector
concerning the index $i$ representing the kind of pulse,
as 
$\overline{\vp}=(\overline{p}_0,\ldots, \overline{p}_{2k})$.

Before the quantum communication, they 
check the probability $p_D$ of dark counts in the detector,
and the probability $p_S$ of errors of the $\times$ basis 
occurred in the detector or the generator,
which can be measured by the error probability when 
the quantum communication channel has no error.
Similarly, they 
the probability $\tilde{p}_S$ of errors of the $+$ basis 
occurred in the detector or the generator.
\begin{enumerate}
\item 
Alice sends the $N'$ pulses, where 
each pulse is chosen among $2k+1$ kinds of pulses.
She denotes the number of the $i$-th kind of pulses by $A_i$.
($\sum_{i=0}^{2k+1}A_i= N'$)
\item 
After sending $N'$ pulses,
Alice announces 
the kind of the each pulse
(the basis and the distribution $\nu_i$)
by using public channel.

\item 
Bob records the numbers $C_0, \ldots, C_{2k}$ of detected pulses and 
the numbers $E_1,\ldots, E_{2k}$ 
of detected pulses with the common basis 
for each kind $i=0,\ldots, 2k$ of pulses.
Bob announces
the positions of pulses with the common basis 
and the above numbers
by using public channel.

\item 
Alice chooses $E_{i_0}-N$ bits among $i_0$-th pulses
with the common basis
and $E_{i_0+k}-N$ bits among the $i_0+k$-th pulses
with the common basis,
and announces these positions and their bit 
by using public channel.
Bob records the number of errors as $H_{i_0}$, $H_{i_0+k}$
and announces them by using public channel.
If $E_{i_0} \le N$ or $E_{i_0+k} \le N$,
they stop their protocol and return to the first step.

\item
Alice and Bob announce their bit of the remaining kinds
$i\neq0,i_0,i_0+k$ of pulses,
and record the number of error by $H_i$.

\item 
Using these informations,
they decide the rates 
$\eta(\frac{H_{i_0+k}}{E_{i_0+k}-N})$
and $\eta(\frac{H_{i_0}}{E_{i_0}-N})$ 
of error correction
and the sizes of
sacrifice bits
$m(\cD_i,\cD_e)$ and
$\tilde{m}(\tilde{\cD}_i,\cD_e)$
in the privacy amplification 
for the remaining the $i_0$-th kind of pulses and 
and the $i_0+k$-th kind of pulses,
respectively,
where
we abbreviate 
the initial data 
$(\vA,\vnu,p_S,p_D)$
and 
$(\vA,\vnu,\tilde{p}_S,p_D)$
and 
the experimental data 
$(\vC,\vE,\vH)$ to 
$\cD_i$, $\tilde{\cD}_i$, and
$\cD_e$.
If $N\eta(\frac{H_{i_0+k}}{E_{i_0+k}-N})
-m(\cD_i,\cD_e)< \underline{N}$ 
or 
$N\eta(\frac{H_{i_0}}{E_{i_0}-N})-
\tilde{m}(\tilde{\cD}_i,\cD_e)<\underline{N}$,
they stop their protocol and return to the first step.
Further, 
if $\overline{N}<
N\eta(\frac{H_{i_0+k}}{E_{i_0+k}-N})-m(\cD_i,\cD_e)$,
they replace $m(\cD_i,\cD_e)$ 
by $N\eta(\frac{H_{i_0+k}}{E_{i_0+k}-N})-\overline{N}$.
Similarly,
if $\overline{N}<
N\eta(\frac{H_{i_0}}{E_{i_0}-N})-\tilde{m}(\tilde{\cD}_i,\cD_e)$,
they replace $\tilde{m}(\tilde{\cD}_i,\cD_e)$.
by
$N\eta(\frac{H_{i_0}}{E_{i_0}-N})-\overline{N}$.

\item
They perform $N$ bits error correction for $+$ basis,
and generate 
$N \eta(\frac{H_{i_0+k}}{E_{i_0+k}-N})$ bits.

\item
They perform privacy amplification for the $+$ basis,
and generate
$N \eta(\frac{H_{i_0+k}}{E_{i_0+k}-N})-
m(\cD_i,\cD_e)$ bits.

\item
They perform $N$ bits error correction for the $\times$ basis,
and generate 
$N \eta(\frac{H_{i_0}}{E_{i_0}-N})$ bits.

\item
They perform privacy amplification for the $\times$ basis,
and generate
$N \eta(\frac{H_{i_0}}{E_{i_0}-N})-
\tilde{m}
(\tilde{\cD}_i,\cD_e)$ 
bits.

\end{enumerate}
If Bob detects
both events $|0\rangle$ and $|1\rangle$
in the measurement of the $+$ basis,
he decides one event with the probability $\frac{1}{2}$.
In the following, this measurement is described by 
the POVM $\{M_{\emptyset},M_0,M_1\}$.

\begin{description}
\item[Error correction (7., 9.)]
In the step 7. and 9.,
Alice and Bob generate
$l+m$ bits 
with negligible errors
from $N$ bits $X$ and $X'$ by using 
one of the following protocols:
(For example, $l$ and $m$ are choosen as
$l+m= N \eta(\frac{H_{i_0+k}}{E_{i_0+k}-N})$ and $m=m(\cD_i,\cD_e)$.)
\begin{description}
\item[Forward error correction]
They share $N \times (l+m)$ binary matrix $M_e$.
Alice generates other $l+m$ bits random number $Z$,
and sends $M_e Z+X$ to Bob.
Bob applies the decoding of the code $M_e$ to the bits
$M_e Z+X-X'$ to extract $Z$, and obtain $Z'$.

\item[Reverse error correction]
Bob generates other $l+m$ bits random number $Z$,
and sends $M_e Z+'X$ to Alice.
Alice applies the decoding of the code $M_e$ to the bits
$M_e Z+X'-X$ to extract $Z$, and obtain $Z'$.

\end{description}
As mentioned later, 
since this error correction corresponds to a part of 
the twirling operation,
their channel can be regarded as
a Pauli channel from Alice to Bob in the forward case
(from Bob to Alice in the reverse case).

\item[Privacy amplification (8., 10.)]
In the step 8. and 10.,
Alice and Bob generate
$l$ bits from $l+m$ bits $Z$ by using 
the following protocol.
First, they generate the same 
$l\times (l+m)$ binary matrix $M_p$ 
with the following condition:
\begin{align}
\rP \{Z \in \im M_p^T\}\le 2^{-m} \label{2-9-1}
\end{align}
for any non-zero $l+m$ bit sequence $Z$.
Next, they generate $l$ bits $M_p Z$ from $l+m$ bits $Z$.
\end{description}

Hence, combining the above error correction and the above 
privacy amplification,
Alice can be regarded to send information by the code
$\im M_e/M_e (\Ker M_p)$.

The preceding researches\cite{mayer1,ILM,WMU}
analyze the security 
when the binary matrix $M_p$ 
for privacy amplification 
is chosen completely randomly.
If we choose 
the binary matrix $M_p$ by the 
Toeplitz matrix\cite{Carter,Krawczyk},
we need less random number.
This is because Toeplitz matrix requires only $l+m-1$ bits random number
while completely random binary matrix $M_p$ does 
$(l+m)l$ bits random number.
An $l\times (l+m)$ binary matrix $(\bX,I)$ 
is called Toeplitz matrix\cite{Carter,Krawczyk}
when its element $\bX=(X_{i,j})$ is given 
by $l+m-1$ random variables $Y_1, \ldots, Y_{l+m-1}$ as 
\begin{align*}
X_{i,j}:=Y_{i+j-1} .
\end{align*}

\begin{thm}\label{th1}
Toeplitz matrix satisfies the condition (\ref{2-9-1})
for any element $Z \neq 0\in \bF_2^{l+m}$.
\end{thm}
For a proof, see Appendix \ref{stoep}.

\section{Evaluation of Eve's information concerning final key}\label{s3}
\subsection{Formulation of channel}
In this section, we assume a simplified Eve's attack,
and evaluate the security against Eve's attack.
In Section \ref{s5}, we will treat the general case of Eve's attack,
and prove that the general case can be reduced to the case of this section.

First, we assume that Eve can distinguish 
the four states $|n,0\rangle$, and $|0,n\rangle$ in $+$ basis and 
$|n,0,\times\rangle$,$|0,n,\times\rangle$ in $\times$ basis.
Hence, the input system can be described by 
$N$-th tensor product system $\cH^{\otimes N}$ of
$\cH:= \cH_0 \oplus \cH_1 \oplus(\oplus_{n\ge 2} \cH_{n,+})
\oplus(\oplus_{n\ge 2} \cH_{n,\times})$,
where 
$\cH_0$ is the one-dimensional space spanned by $|0,0 \rangle $,
$\cH_1$ is the two-dimensional space spanned by $|0,1 \rangle$ and 
$|1,0 \rangle$,
$\cH_{n,+}$ is
the two-dimensional space spanned by $|n,0\rangle$, and $|0,n\rangle$,
and 
$\cH_{n,\times}$ is
the two-dimensional space spanned by $|n,0,\times\rangle$ and 
$|0,n,\times\rangle$.
The output system is described by 
$N$-th tensor product space $\cH^{\otimes N}$ of $\cH_0 \oplus \cH_1$.

Then, the quantum communication channel from Alice to Bob is given by
\begin{align}
\bigoplus_{\vec{n}}
\sum_{\vec{e}}
\cP_{\vec{n}}(\vec{e})
(\otimes_{i=1}^{N}{\cal E}_{e_i|n_i})
(\rho)\label{1-19-2},
\end{align}
where $\cP_{\vec{n}}(\vec{e})$ is
the distribution of $\vec{e}$ when $\vec{n}$ is fixed.
Such a channel is called Pauli channel.
Here, $\vec{n}$ and $\vec{e}$ are given as follows:
Each element $n_i$ of 
$\vec{n}=(n_1, \ldots, n_N)$ is chosen among 
$0,1,(2,+),(2,\times), \ldots$.
Each element $e_i$ of $\vec{e}=(e_1, \ldots, e_N)$ 
is chosen as $v$ or $s$ when $n_i$ is $0$.
It is chosen among $v,(0,0),(0,1),(1,0),(1,1)$ 
when $n_i$ is $1$.
Otherwise, it is chosen among $v,0,1$.
When $n_i$ is $0$ or $1$,
the channel ${\cal E}_{e_i|n_i}$ is defined as 
\begin{align*}
{\cal E}_{e_i|0}(\rho)
&:= 
\left\{
\begin{array}{ll}
\langle 0,0| \rho| 0,0 \rangle 
| 0,0 \rangle \langle 0,0| 
& \hbox{ if }e_i=v \\
\langle 0,0| \rho| 0,0 \rangle 
\rho_{mix,1}
 & \hbox{ if }e_i=s 
\end{array}
\right. \\
{\cal E}_{e_i|1}(\rho)
&:= 
\left\{
\begin{array}{ll}
| 0,0 \rangle \langle 0,0|  \Tr P_{\cH_1} \rho P_{\cH_1} & 
\hbox{ if }
e_i=v \\
\sW^{e_i} 
P_{\cH_1} \rho P_{\cH_1}
(\sW^{e_i} )^{\dagger}
& \hbox{ otherwise,}
\end{array}
\right. 
\end{align*}
where 
\begin{align*}
\rho_{mix,1}&:=
\frac{1}{2}(| 0,1 \rangle \langle 0,1|+| 1,0 \rangle \langle 1,0| ),\\
\sW^{(x,z)}&:=\sX^x \sZ^z,\\
\sX | 0 ,1\rangle &=| 1, 0\rangle ,~
\sX | 1 ,0\rangle =| 0,1\rangle, \\
\sZ | 0 ,1\rangle &=-| 0,1\rangle ,~
\sZ | 1 ,0\rangle =| 1,0\rangle .
\end{align*}
When $n_i$ is not $0$ or $1$,
the channel ${\cal E}_{e_i|n_i}$ is defined as 
\begin{widetext}
\begin{align*}
{\cal E}_{e_i|n_i}(\rho)
:= 
\left\{
\begin{array}{ll}
| 0,0 \rangle \langle 0,0| \Tr \rho P_{\cH_{n_i}}
 & \hbox{ if }e_i=v \\
\langle 0 ,n_i|\rho| 0 , n_i\rangle
| 0 , n_i\rangle \langle 0 ,n_i|
+
\langle 1,n_i|\rho|1, n_i\rangle
|1 , n_i\rangle \langle 1 ,n_i|
 & \hbox{ if }e_i=0 \\
\langle 0 ,n_i|\rho| 0 , n_i\rangle
| 1 , n_i\rangle \langle 1 ,n_i|
+
\langle 1,n_i|\rho|1, n_i\rangle
|0 , n_i\rangle \langle 0 ,n_i|
 & \hbox{ if }e_i=1.
\end{array}
\right. 
\end{align*}
\end{widetext}

The raw key is generated from 
detected pulses, which belong to the system $\cH_1$ on the Bob's side.
Thus, we focus only on the pulse whose measurement value is not $v$.
In the following, we consider the security from the final key 
distilled from raw keys of the $+$ basis.
Hence, the generated state can be restricted to 
$\cH_0 \oplus \cH_1 \oplus(\oplus_{n\ge 2} \cH_{n,+})$.
Thus, it can be assumed that our channel 
$\bigoplus_{\vec{n}}
\sum_{\vec{e}}
\cP_{\vec{n}}(\vec{e})
(\otimes_{i=1}^{N}{\cal E}_{e_i|n_i})$
satisfies that 
each element $e_i$ of $\vec{e}= (e_1, \ldots, e_{N})$ 
is not $v$.

\subsection{Security of known channel: no dark count case}\label{2-16-1}
Assume that the input state belongs to the subsystem 
$\cH_{\vec{n}}:=\cH_{n_1} \otimes \cdots \otimes \cH_{n_{N}}$ labeled by 
$\vec{n}= (n_1, \ldots, n_{N})$.
Now, we classify the $N$ input subsystems into three parts:
\begin{description}
\item[0th part: ]
$K^0(\vec{n}):= \#\{i| n_i=0\}$.
\item[1st part: ]
$K^1(\vec{n}):= \#\{i| n_i=1\}$.
\item[2nd part: ]
$K^2(\vec{n}):= \#\{i| n_i\ge 2\}$.
\end{description}
In the $0$-th part, 
Eve can obtain no information.
That is, Eve's information 
is equal to Eve's information 
when the Alice's information 
is sent by the $+$ basis
via the qubit channel:
\begin{align*}
{\cal E}_{s|0}'(\rho):= 
\frac{1}{2}
(\sX^0 \rho (\sX^0 )^\dagger
+
\sX^1 \rho (\sX^1 )^\dagger).
\end{align*}
In the 2nd part, 
Eve can obtain all of Alice's information
by the following method:
Eve receives two-photon state.
She sends one qubit system to Bob,
and keeps the other qubit.
After the announcement of the basis,
Eve measures her system with the correct basis.
Thus,
Eve's information 
is equal to Eve's information 
when the Alice's information 
is sent by the $+$ basis
via the phase-damping
qubit channel (pinching channel):
\begin{align*}
{\cal E}_{e_i|(n,+)}'(\rho)
:= 
\frac{1}{2}
(\sZ^0 \rho (\sZ^0 )^\dagger
+
\sZ^1 \rho (\sZ^1 )^\dagger)
\end{align*}
for $n \ge 2$.
This is because 
the channel is given by ${\cal E}_{e_i|(n,+)}'$
in the single photon case
when Eve measures the system with the correct basis.
Here, 
the presence or the absence of the error $\sX$ 
is not so important for Eve's information.
This is because 
the probabilities concerning the action $\sZ$ 
is essential, as is discussed in Appendix \ref{as3}. 
Therefore,
Eve's information concerning total $N$ bits
is equal to 
Eve's information 
when the $N$ bits information $x_1, \ldots, x_{N}$ 
is sent by $|x_1, \ldots, x_{N}\rangle \in (\complex^2)^{\otimes N}$
via the following qubit channel:
\begin{align}
\sum_{\vec{e}}\cP_{\vec{n}}(\vec{e})
(\otimes_{i=1}^N{\cal E}_{e_i|n_i}').\label{3-15-4}
\end{align}
There is a relation between the error probability 
in the $\times$ basis and the security.

\begin{thm}
\label{t-6-1}
Define $P^{\cP_{\vec{n}}}_{ph| M_p}$ 
as the error probability by
an arbitrary decoding
when an information sent by 
the code $(M_e (\Ker M_p))^{\perp} /(\im M_e)^{\perp}$
with the $\times$ basis 
via the qubit channel 
$\sum_{\vec{e}}\cP_{\vec{n}}(\vec{e})
(\otimes_{i=1}^N{\cal E}_{e_i|n_i}')(\rho)$,
where $\vec{n}:=(n_1, \ldots, n_{N})$.
When Alice sends $l$ bits information with the code
$\im M_e/M_e (\Ker M_p)$
in the $+$ basis via the same channel, 
the following relations hold.

Define $\rho^E_{[Z]|M_p}$ as the final Eve's state
when Alice's information is $[Z]= M_p Z$.
Then, Eve's information 
$I^{\cP_{\vec{n}}}_{E|M_p}$ is given as
the quantum mutual information:
\begin{align*}
I^{\cP_{\vec{n}}}_{E|M_p}&:=
\frac{1}{2^l}
\sum_{[Z]}
D(\rho^E_{[Z]|M_p}\|\overline{\rho}^E_{M_p})\\
D(\rho\|\rho')&:=
\Tr \rho (\log \rho - \log \rho')\\
\overline{\rho}^E_{M_p}
&:= \sum_{[Z]}\frac{1}{2^l}\rho^E_{[Z]|M_p}.
\end{align*}
The quantum mutual information 
$I^{\cP_{\vec{n}}}_{E|M_p}$ satisfies that 
\begin{align}
I^{\cP_{\vec{n}}}_{E|M_p}\le 
\overline{h}(P^{\cP_{\vec{n}}}_{ph| M_p})
+ 
l P^{\cP_{\vec{n}}}_{ph| M_p}\label{1-19-6-2},
\end{align}
where $\overline{h}(x)$ is defined as
\begin{align*}
\overline{h}(x):=
\left\{
\begin{array}{ll}
-x \log_2 x -(1-x)\log_2 (1-x) & \hbox{ if } 0 \le x \le 1/2\\
1 & \hbox{ if } 1/2 < x \le 1.
\end{array}
\right.
\end{align*}
Hence, Eve's information per one bit is evaluated 
as
\begin{align}
\frac{I^{\cP_{\vec{n}}}_{E|M_p}}{l}
\le 
\frac{\overline{h}(P^{\cP_{\vec{n}}}_{ph| M_p})}{l}
+ 
P^{\cP_{\vec{n}}}_{ph| M_p}\label{1-19-7-2}.
\end{align}
We also obtain the following:
\begin{align}
\min_{[Z]\neq[Z']}
F(\rho^E_{[Z]| M_p},\rho^E_{[Z']| M_p})
& \ge
1-2 P^{\cP_{\vec{n}}}_{ph| M_p}\label{1-19-8}\\
\max_{[Z]\neq[Z']}
\|\rho^E_{[Z]| M_p}-\rho^E_{[Z']| M_p}
\|_1
&\le
4 P^{\cP_{\vec{n}}}_{ph| M_p}.\label{1-19-9}\\
\min_{[Z]}
F(\rho^E_{[Z]| M_p},\overline{\rho}^E_{M_p})
& \ge
1- P^{\cP_{\vec{n}}}_{ph| M_p}\label{1-19-8-2}\\
\max_{[Z]}
\|\rho^E_{[Z]| M_p}-\overline{\rho}^E_{M_p}\|_1
&\le
2 P^{\cP_{\vec{n}}}_{ph| M_p},\label{1-19-9-2}
\end{align}
where
$F(\rho,\rho'):= \Tr \sqrt{\sqrt{\rho'}\rho\sqrt{\rho'}}$.
Define $P^{\cP_{\vec{n}}}_{succ|M_p}$ as the probability 
of successfully detecting the Alice's information $[Z]$.
Then, the inequality
\begin{align}
P^{\cP_{\vec{n}}}_{succ|M_p}
\le 
\left(
\sqrt{P^{\cP_{\vec{n}}}_{ph| M_p}}
\sqrt{1-2^{-l}}
+
\sqrt{1-P^{\cP_{\vec{n}}}_{ph| M_p}}
\sqrt{2^{-l}}
\right)^2\label{1-31-1}
\end{align}
holds.

Further, the concavity of $\overline{h}$ implies that
\begin{align}
&\rE^{\cP_{\vec{n}}}_{M_p}
I^{\cP_{\vec{n}}}_{E|M_p}\le 
\overline{h}(\rE^{\cP_{\vec{n}}}_{M_p} P^{\cP_{\vec{n}}}_{ph| M_p})
+ 
l\rE^{\cP_{\vec{n}}}_{M_p} 
P^{\cP_{\vec{n}}}_{ph| M_p}\label{1-19-6}\\
&\rE^{\cP_{\vec{n}}}_{M_p}
\frac{I^{\cP_{\vec{n}}}_{E|M_p}}{l}
\le 
\frac{\overline{h}(\rE^{\cP_{\vec{n}}}_{M_p} P^{\cP_{\vec{n}}}_{ph| M_p})
}{l}
+ 
\rE^{\cP_{\vec{n}}}_{M_p} P^{\cP_{\vec{n}}}_{ph| M_p}\label{1-19-7}.
\end{align}
The concavity of left hand side of 
(\ref{1-31-1}) holds concerning $P^{\cP_{\vec{n}}}_{ph| M_p}$.
Thus, 
\begin{align*}
&\rE^{\cP_{\vec{n}}}_{M_p}P^{\cP_{\vec{n}}}_{succ|M_p}\\
\le &
\left(
\sqrt{\rE^{\cP_{\vec{n}}}_{M_p}P^{\cP_{\vec{n}}}_{ph| M_p}}
\sqrt{1-2^{-l}}
+
\sqrt{1-\rE^{\cP_{\vec{n}}}_{M_p}P^{\cP_{\vec{n}}}_{ph| M_p}}
\sqrt{2^{-l}}
\right)^2.
\end{align*}
\end{thm}
For a proof, see Appendix \ref{as3}.
As shown in Section \ref{s5}, 
sending $l$ bits information with the code
$\im M_e/M_e (\Ker M_p)$
is equivalent with 
the combination of 
sending random number and 
forward error correction by $\im M_e$ and privacy amplification by 
$M_e (\Ker M_p)$.

Next, we focus on 
the average error probability 
$P^{\cP_{\vec{n}}}_{ph, min| M_p}$ 
with the minimum length decoding
when an information sent by 
the code $(M_e (\Ker M_p))^{\perp}/(\im M_e)^{\perp}$
with the $\times$ basis 
via the qubit channel 
$\Lambda(\rho)=
\sum_{\vec{e}}\cP_{\vec{n}}(\vec{e})
(\otimes_{i=1}^N{\cal E}_{e_i|n_i}')(\rho)$.
This value is described as
\begin{align*}
P^{\cP_{\vec{n}}}_{ph, min| M_p}
=
\frac{1}{|(\im M_e)^{\perp}|}
\sum_{z\in (\im M_e)^{\perp}}
\sum_{z': {(\ref{2-15-1})}}
\langle z'|
\cP_{\vec{n}}(|z \rangle \langle z|)
|z'\rangle,
\end{align*}
where we take the summand concerning $z'$ satisfying the following condition 
(\ref{2-15-1}):
\begin{align}
\argmin_{z''\in 
(M_e (\Ker M_p))^{\perp}
}\dis(z',z'')
\in (\im M_e )^{\perp},
\label{2-15-1}
\end{align}
and $\dis(z',z'')$ is the Hamming distance between $z'$ and $z''$.
In order to analyze the error probability
$P^{\cP_{\vec{n}}}_{ph, min| M_p}$,
we introduce the number $t(\vec{e},\vec{n})$:
\begin{align*}
t(\vec{e},\vec{n}):= \#\{
i| n_i=1,~ e_i= (0,1) \hbox{ or } (1,1)\}.
\end{align*}

\begin{thm}\label{t-6-2}
Assume that the binary matrix $M_p$ satisfies the condition (\ref{2-9-1}).
If the distribution $\cP_{\vec{n}}$ takes positive probabilities 
only in the set $\{\vec{e}|t(\vec{e},\vec{n})=t\}$,
then 
we obtain 
\begin{align*}
\rE_{M_p} P^{\cP_{\vec{n}}}_{ph, min| M_p}
\le 2^{K^1(\vec{n}) \overline{h}(
\frac{t}{K^1(\vec{n})}
)+K^2(\vec{n})-m}.
\end{align*}
Further, 
if the stochastic behavior of 
the random variable $t=t(\vec{e},\vec{n})$ on 
the distribution $\cP_{\vec{n}}$
is described by the distribution
$p(\frac{t}{K^1(\vec{n})})$,
then the inequality
\begin{align*}
&\rE_{M_p} P^{\cP_{\vec{n}}}_{ph, min| M_p}\\
\le &
\sum_{t=0}^{K^1(\vec{n})}
p(\frac{t}{K^1(\vec{n})})
\min\left\{
2^{K^1(\vec{n}) \overline{h}(\frac{k}{K^1(\vec{n})})+K^2(\vec{n})-m},1\right\}
\end{align*}
holds.
That is, the upper bound can be characterized by 
$\vec{K}(\vec{n})$ and $t$.
\end{thm}
For a proof, see Appendix \ref{as3-2}.

\subsection{Security of known channel: dark count case}
Next, we take into account the effect of dark count in the detector.
In this case, in order to characterize
the presence or the absence of dark count,
we add $c$ or $d$ to the label $n_i$ of the input system.
That is, the label $n_i$ is chosen among 
$(0,c),(0,d),(1,c),(1,d),(2,+,c),(2,+,d),(2,\times,c),(2,\times,d),
\ldots$ etc, where $(*,d)$ expresses dark count
and $(*,c)$ does the normal count.
Then, we can classify 
detected pulses to the following six parts:
\begin{description}
\item[$o=0$, $J^0(\vec{n})$:] 
The number of detected pulses except for dark count
whose initial (Alice's) state is the vacuum state.
\item[$o=1$, $J^1(\vec{n})$:] 
The number of detected pulses except for dark count
whose initial (Alice's) state is the single-photon state.
\item[$o=2$, $J^2(\vec{n})$:]
The number of detected pulses except for dark count
whose initial (Alice's) state is the multi-photon state.
\item[$o=3$, $J^3(\vec{n})$:]
The number of pulses detected by dark count
whose initial (Alice's) state is the vacuum state.
\item[$o=4$, $J^4(\vec{n})$:]
The number of pulses detected by dark count
whose initial (Alice's) state is the single-photon state.
\item[$o=5$, $J^5(\vec{n})$:] 
The number of detected pulses except for dark count
whose initial (Alice's) state is the multi-photon state.
\end{description}

Now, we consider the following protocol:
First, Alice sends the random number with 
the $+$ basis via 
$\sum_{\vec{e}}\cP_{\vec{n}}(\vec{e})
(\otimes_{i=1}^N{\cal E}_{e_i|n_i})(\rho)$,
where 
for the dark counts $n_i=(*,d)$,
$e_i$ takes only $d$ and 
the map ${\cal E}_{d|n_i}$ is given by
$\cE_{d|n_i}(\rho)= \frac{1}{2}(
|0,1\rangle \langle 0,1|+
|1,0\rangle \langle 1,0|)$.
Second, they apply the forward or reverse error correction 
by the code $\im M_e$,
and finally perform privacy amplification by $M_p$,
where $M_p$ is assumed to satisfy (\ref{2-9-1}).
In this case, we obtain the same argument as Theorem \ref{t-6-1}.
Thus, in order to discuss the security, we need to characterize 
the average error probability in the $\times$ basis. 

Now, we consider the forward error correction case.
In the event $o=0,3$, Eve cannot obtain any information of Alice's raw key.
Also,  
in the event $o=2,4,5$,
Eve can obtain all information of Alice's raw key.
Thus, our situation is the same as
the case of $K^1= J^1$ and
$K^2= J^2+J^4+J^5$ of Theorem \ref{t-6-2}.
Similar to the above subsection,
we define
the average error probability 
$P^{\cP_{\vec{n}}}_{ph, min,\to| M_p}$
of the code $(M_e (\Ker M_p))^{\perp} /(\im M_e)^{\perp}$
concerning the $\times$ basis with the channel 
$\sum_{\vec{e}}\cP_{\vec{n}}(\vec{e})
(\otimes_{i=1}^N{\cal E}_{e_i|n_i}')(\rho)$,
where we define 
the map ${\cal E}_{e_i|n_i}'$ for dark count $n_i=(*,d)$ as follows:
${\cal E}_{d|(*,d)}'$ is the same as
${\cal E}_{d|*}$ and 
\begin{align*}
{\cal E}_{d|(0,d)}'(\rho)
&:= 
\frac{1}{2}(
\sX^{0}\rho(\sX^{0})^{\dagger}
+
\sX^{1}\rho(\sX^{1})^{\dagger}),\\
{\cal E}_{d|(*,d)}'(\rho)
&:= 
\frac{1}{2}(
\sZ^{0}\rho(\sZ^{0})^{\dagger}
+
\sZ^{1}\rho(\sZ^{1})^{\dagger})
\end{align*}
for $* \neq 0$.
The distribution $p(\frac{t}{J^1(\vec{n})})$
is defined as the distribution describing 
the random variable $t=t(\vec{e},\vec{n})$ 
under the distribution $\cP_{\vec{n}}$,
and define $t(\vec{e},\vec{n})$ by
\begin{align*}
t(\vec{e},\vec{n}):= \#\{
i| n_i=(1,c),~ e_i= (0,1) \hbox{ or } (1,1)\}.
\end{align*}
Then, we obtain 
\begin{align}
&\rE_{M_p} P^{\cP_{\vec{n}}}_{ph, min,\to | M_p} \nonumber\\
\le &
\sum_{t=0}^{J^1}
p(\frac{t}{J^1})
\min\left\{
2^{J^1 \overline{h}(\frac{t}{J^1})+J^2+J^4+J^5-m},1\right\}
.\label{2-16-2}
\end{align}

Next, we consider the reverse error correction case.
We assume that the bits detected by dark count cannot be 
controlled by Eve.
That is, in the event $o=3,4,5$, 
Eve cannot obtain any information of Bob's raw key.
Also,  
in the event $o=0,2$,
Eve can obtain all information of Bob's raw key.
(In the case of $o=0$,
Eve can obtain Bob's information
by the following.
Eve generates an entangled pair, and sends Bob a part of it.
After announcing the basis, Eve measures the remaining part based on the 
correct basis.)
Hence,
our situation is the same as
the case of $K^1= J^1$ and
$K^2= J^0+J^2$ of Theorem \ref{t-6-2}.
Similar to the above subsection,
we define
the average error probability 
$P^{\cP_{\vec{n}}}_{ph, min,\leftarrow | M_p}$
of the code $(M_e (\Ker M_p))^{\perp} /(\im M_e)^{\perp}$
concerning the $\times$ basis with the channel 
$\sum_{\vec{e}}\cP_{\vec{n}}(\vec{e})
(\otimes_{i=1}^N{\cal E}_{e_i|n_i}')$,
where we define 
the map ${\cal E}_{e_i|n_i}'$ as follows:
\begin{align*}
{\cal E}_{d|(\cdot,d)}'(\rho)
&:=
\frac{1}{2}(
\sX^{0}\rho(\sX^{0})^{\dagger}
+
\sX^{1}\rho(\sX^{1})^{\dagger}),\\
{\cal E}_{e_i|(1,c)}'(\rho)
&:= 
\sW^{e_i} \rho (\sW^{e_i} )^{\dagger},\\
{\cal E}_{e_i|(*,c)}'(\rho)
&:= 
\frac{1}{2}(
\sZ^{0}\rho(\sZ^{0})^{\dagger}
+
\sZ^{1}\rho(\sZ^{1})^{\dagger})
\end{align*}
for $* \neq 1$.

Note that the definition of ${\cal E}_{e_i|n_i}'$ 
for dark count $n_i=(*,d)$ 
is different from the forward case.
Here $x=\leftarrow$ expresses the reverse case.
Then, we obtain 
\begin{align}
&\rE_{M_p} P^{\cP_{\vec{n}}}_{ph, min,\leftarrow | M_p}\nonumber\\
\le &
\sum_{t=0}^{J^1}
p(\frac{t}{J^1})
\min\left\{
2^{J^1 \overline{h}(\frac{t}{J^1})+J^0+J^2-m},1\right\}.
\label{2-16-3}
\end{align}

\subsection{Security of unknown channel: dark count case}\label{s3d}
Now, we back to the original setting.
Since the numbers $J^0,\ldots, J^5$ and the ratio $r^1
:= \frac{t}{J^1}$ are unknown,
the size of sacrifice bits
is chosen as the function 
$m(\cD_i,\cD_e)$ 
of the random variable $\cD_e$.
For simplicity, 
we abbreviate
$m(\cD_i,\cD_e)$ and 
$\eta(\frac{H_{i_0+k}}{E_{i_0+k}-N})$
to $m$ and $\eta$.

Now, we give general security formulas 
for the given function $m$ of $\cD_e$.
The random variable $\vec{n}$ is known by Eve, 
but cannot be decided by Eve.
Hence, Eve's information is measured by 
the conditional expectations 
$I^{\cP}_{E|M_p,\cD_e,\POS}$ 
of $I^{\cP_{\vec{n}}}_{E|M_p}$
concerning the random variable $\vec{n}$
when the random variables $M_p,\cD_e$, and $\POS$
are fixed,
where $\POS$ is the random variable describing the position of
the check bits and each kinds of pulses.
We define
$P^{\cP}_{ph,min,x|M_p,\cD_e,\POS}$ 
as the conditional expectations of 
$P^{\cP_{\vec{n}}}_{ph, min,x| M_p,\vec{J}}$
concerning $\vec{n}$ 
when the random variables $\cD_e$, and $\POS$
are fixed.
Then, we obtain 
\begin{align}
&\rE^{\cP}_{M_p,\cD_e,\POS} 
I^{\cP}_{E|M_p,\cD_e,\POS}\nonumber\\
\le & 
\rE^{\cP}_{\cD_e,\POS} 
\overline{h}(P^{\cP}_{ph,min,x|\cD_e,\POS})\nonumber\\
&\quad + 
\rE^{\cP}_{\cD_e,\POS} 
(
N\eta-m) 
P^{\cP}_{ph,min,x|\cD_e,\POS}\nonumber\\
\le &
P^{\cP}_{ph,av,x}
(\overline{N}+1-\log P^{\cP}_{ph,av,x})\label{2-9-2},
\end{align}
where
$\rE^{\cP}_{M_p,\cD_e,\POS}$
($\rE^{\cP}_{\cD_e,\POS}$) 
denotes the 
expectation concerning 
the random variables $M_p,\cD_e$, and $\POS$,
($\cD_e$, and $\POS$).
The inequality (\ref{2-9-2}) is proved in Appendix \ref{2-14-1}.
Hence, Eve's information per one bit can be evaluated as
\begin{align*}
 \rE^{\cP}_{\cD_e,\POS} 
\frac{I^{\cP}_{E|\cD_e,\POS}}{
N\eta-m}
\le &
\rE^{\cP}_{\cD_e,\POS} 
\frac{\overline{h}(
P^{\cP}_{ph|\cD_e,\POS})}{
N\eta-m}
+ 
P^{\cP}_{ph,av,x}\\
\le &
\frac{\overline{h}(
P^{\cP}_{ph,av,x}
)}{
\underline{N}}
+ 
P^{\cP}_{ph,av,x}.
\end{align*}
Similarly, Eve's state can be given as
the conditional average Eve's state 
${\rho}^E_{[Z]|M_p,\cD_e,\POS}$ 
with the final key $[Z]$
when the random variables $M_p,\cD_e$, and $\POS$
are fixed.
Then,
\begin{align}
&\rE^{\cP}_{M_p,\cD_e,\POS} 
\min_{[Z]\neq[Z']}
F({\rho}^E_{[Z]},{\rho}^E_{[Z']|M_p,\cD_e,\POS})\nonumber\\
 \ge &
1-2 P^{\cP}_{ph,av,x}\nonumber\\
&\rE^{\cP}_{M_p,\cD_e,\POS} 
\max_{[Z]\neq[Z']}
\|{\rho}^E_{[Z]|M_p,\cD_e,\POS}-
{\rho}^E_{[Z']|M_p,\cD_e,\POS}\|_1\nonumber\\
\le &
4 P^{\cP}_{ph,av,x}\nonumber\\
& \rE^{\cP}_{M_p,\cD_e,\POS} 
\min_{[Z]}
F({\rho}^E_{[Z]|M_p,\cD_e,\POS},
\overline{\rho}^E_{M_p,\cD_e,\POS})\nonumber\\
\ge &
1- P^{\cP}_{ph,av,x}\nonumber\\
& \rE^{\cP}_{M_p,\cD_e,\POS} 
\max_{[Z]}
\|{\rho}^E_{[Z]|M_p,\cD_e,\POS}
-\overline{\rho}^E_{M_p,\cD_e,\POS}\|_1 \nonumber\\
\le &
2 P^{\cP}_{ph,av,x},\nonumber
\end{align}
where
$\overline{\rho}^E_{M_p,\cD_e,\POS}$ 
is the average state of
${\rho}^E_{[Z]|M_p,\cD_e,\POS}$ concerning $[Z]$,
and $P^{\cP}_{ph,av,x}:=
\rE^{\cP}_{\cD_e,\POS} 
P^{\cP}_{ph|\cD_e,\POS}$.
We can evaluate the probability 
$P^{\cP}_{succ,x|M_p,\cD_e,\POS}$ 
that Eve successfully detects the final key $[Z]$:
\begin{align}
& \rE^{\cP}_{M_p,\cD_e,\POS} 
P^{\cP}_{succ,x|M_p}\nonumber \\
\le &
\rE^{\cP}_{M_p,\cD_e,\POS} 
\biggl(
\sqrt{
P^{\cP}_{ph,min,x|M_p,\cD_e,\POS}
}
\sqrt{1-2^{-(\eta N-m)}} \nonumber\\
&\quad +
\sqrt{1-P^{\cP}_{ph,min,x|M_p,\cD_e,\POS}}
\sqrt{2^{-(\eta N-m)}}
\biggr)^2\label{1-31-1-3}\\
\le &
\left(
\sqrt{P^{\cP}_{ph,av,x}}
\sqrt{1-2^{-\underline{N}
}}
+
\sqrt{1-P^{\cP}_{ph,av,x}}
\sqrt{2^{-\underline{N}}}
\right)^2\label{1-31-1-2},
\end{align}
where (\ref{1-31-1-2}) follows from
the concavity of left hand side of 
(\ref{1-31-1}).

In order to guarantee the security,
it is sufficient to show that
the probability $P^{\cP}_{ph,av,x}$ 
is quite small for any $\cP$.
Since the quantity $P^{\cP}_{ph,av,x}$ 
has the linear form concerning $\cP$,
it is enough to treat $P^{\cP}_{ph,av,x}$ when 
$\cP$ is an extremal point.
That is, the relation 
\begin{align*}
&\max_{\cP:\hbox{ any conditional distribution}}
P^{\cP}_{ph,av,x}\\
=&
\max_{\cP \in {\cal EP}}
P^{\cP}_{ph,av,x}
\end{align*}
holds,
where ${\cal EP}$ is the set of extremal points
concerning the set of conditional distributions.
From (\ref{2-16-2}) and (\ref{2-16-3}),
these values are evaluated as follows.
\begin{align}
& 
\max_{\cP \in {\cal EP}}
P^{\cP}_{ph,av,\to} \nonumber\\
\le &
\max_{\cP \in {\cal EP}}
\rE^{\cP}_{\vec{J},t,\cD_e,\POS} 
\!\!\min\left\{2^{J^1 \overline{h}(\frac{t}{J^1})+J^2+J^4+J^5-m},
1\right\}\label{1-29-2}\\
& 
\max_{\cP \in {\cal EP}}
P^{\cP}_{ph,av,\leftarrow }\nonumber\\
\le &
\max_{\cP \in {\cal EP}}
\rE^{\cP}_{\vec{J},t,\cD_e,\POS} 
\!\!
\min\left\{2^{J^1 \overline{h}(\frac{t}{J^1})+J^0+J^2-m},1 \right\}.
\label{1-29-1}
\end{align}
where $t$ is the number of errors of the $\times$ basis
in the event of $o=1$. 
Here, we have to treat the conditional expectation
concerning $\vec{J}$ and $t$ 
even if the other random variable $\cD_e$ 
is fixed.
Hence, our purpose is choosing 
the size $m$ of sacrifice bits based on 
the information $\cD_i$ and $\cD_e$.

Since Alice chooses
the positions of  each kinds of pulses 
and check bits randomly,
as is discussed in Hayashi et al. \cite{H3}, 
the stochastic behavior of $\cD_e$
is given by hypergeometric distribution 
in the case of any extremal point $\cP$.
In order to guarantee the security with the finite-length code,
we have to calculate 
(\ref{1-29-2}) and (\ref{1-29-1}) for 
the specific function 
$m(\cD_i,\cD_e)$.
Since this task needs a large amount of calculation 
due to a large number of random variables,
we treat it in another paper \cite{H3}.

In the beginning of this section,
we assume that 
the states 
$\{|n,0\rangle\langle n,0|,|0,n\rangle\langle 0,n|,
|n,0,\times\rangle\langle n,0,\times|,
|0,n,\times\rangle\langle 0,n,\times|\}_{n\ge 2}$
can be distinguished by Eve.
Now, instead of the above states,
we focus on the other set of states
$\{
\rho_{i}^{\uparrow,+}:=
\sum_n s^i_n
|n,0\rangle\langle n,0|
,
\rho_{i}^{\downarrow,+}:=
\sum_n s^i_n
|0,n\rangle\langle 0,n|,
\rho_{i}^{\uparrow,\times}:=
\sum_n s^i_n
|n,0,\times\rangle\langle n,0,\times|,
\rho_{i}^{\downarrow,\times}:=
\sum_n s^i_n
|0,n,\times\rangle\langle 0,n,\times|\}_{n\ge 2}$
which can describe all sent pulses by the convex combination of 
theirselves 
with the states
$|0,0\rangle\langle 0,0| $,
$|1,0\rangle\langle 1,0| $,
$|0,1\rangle\langle 0,1| $,
$|1,0,\times\rangle\langle 1,0,\times| $,
$|0,1,\times\rangle\langle 0,1,\times| $.
Then, we can assume so stronger ability of Eve that
Eve can distinguish all states of 
$\{
\rho_{i}^{\uparrow,+},
\rho_{i}^{\downarrow,+},
\rho_{i}^{\uparrow,\times},
\rho_{i}^{\downarrow,\times}\}_{n\ge 2}$.
In this case, we obtain the same argument as this section 
with replacing the former set by the latter set.
The construction of $s^i_n$ 
in the case of the phase-randomized coherent light 
is given in Hayashi \cite{H2}.

\subsection{Security with two-way error correction}\label{s3e}
Here, we should remark that 
the effects of dark counts and the vacuum states
are helpful only when 
the error correction is one-way.
If we apply a careless two-way error correction,
these effects are not so helpful.
That is, 
Eve has a possibility to access the information in the events $o=0,2,3,4,5$.
The main point of the two-way error correction
is the following:
Consider the case where 
a reverse error correction is applied after 
a forward error correction.
In this case, the second error correction depends on
(a prat of) Bob's syndrome.
That is, he has to announce (a part of) his syndrome.
Now, consider an extremal case, i.e., the case where
Bob announces all of his syndrome.
This case is equivalent with the case where
Bob announces his syndrome after Alice transmits 
her information via a Pauli channel with the $+$ basis.

In the single-photon case,
as is discussed in Appendix \ref{a2},
Eve's information contains 
all information concerning the flip action $\sX$ on the $+$ basis,
which includes Bob's syndrome.
Hence, this information it is useless for Eve 
in the single-photon case.
However, it allows Eve to access the information 
in the events $o=0,2,4,5$ in the imperfect photon case.
Eve knows the parts $o=2,4,5$ concerning Alice's bits $Z$
after the forward error correction by the code $C \subset \bF_2^N$.
She also knows the parts $o=0,2$ concerning Bob's bits $Z'$
after the forward error correction by the code $C$
using Bob's syndrome.
The channels in other parts $o=1,3$ can be
regarded as the single-photon case with the channels:
\begin{align*}
\cE_{e_i|(1,c)}'(\rho)&= \sW^{e_i}\rho(\sW^{e_i})^{\dagger}\\
\cE_{d|(0,d)}'(\rho)&= 
\frac{1}{2}(
\sZ^{0}\rho(\sZ^{0})^{\dagger}
+
\sZ^{1}\rho(\sZ^{1})^{\dagger}).
\end{align*}
Suppose that Bob can perfectly correct the error,
i.e., his bits $Z'$ is equal to hers $Z$.
Eve knows the parts $o=0,2,4,5$ concerning $Z$.
Now, we focus on 
the subcode $C'\subset \bF_2^{J^1+J^3}$
defined by
\begin{align*}
C'=\{ x\in \bF_2^{J^1+J^3}| 
(x,\vec{0}_{N-(J^1+J^3)})\in C \},
\end{align*}
where $\vec{0}_{N-(J^1+J^3)}$ is the $0$ vector in 
the composite system of the parts $o=0,2,4,5$.
Then, Eve's state is equal to that in the case where
Alice sends her information with the code $C'$ via the 
$J^1+J^3$-qubits channel
$\sum_{e_i}
P_{\vec{n}}(\vec{e})
\otimes_{i:n_i=(1,c),(0,d)}\cE_{e_i|n_i}'$.
Therefore,
our situation is the same as
the case of $K^1= J^1$ and
$K^2= J^0+J^2+J^4+J^5$ of Theorem \ref{t-6-2}.

Now, we proceed to the general case of two-way error correction,
in which 
the final classical error correction code $C^u$ is chosen with the probability
$p(u)$, i.e.,
Alice decides the $i$-th code $C_i$ depending on the $i-1$ syndromes of Bob,
inductively.
We define the average error probability 
$P^{\cP_{\vec{n}}}_{ph, min,\leftrightarrow| M_p,u}$
and
the distribution $p_u(\frac{t}{J^1})$ concerning 
the random variable $\frac{t}{J^1}$ when the classical error 
correction code $C^u$ is chosen,
where $t$ is defined similar to subsection \ref{2-16-1}.
The relation
\begin{align}
\sum_{u}p(u)
p_u(\frac{t}{J^1})=p(\frac{t}{J^1})\label{2-23-1}
\end{align}
holds.
Applying Theorem \ref{t-6-2} in the case of
$K^1= J^1$ and $K^2= J^0+J^2+J^4+J^5$,
we obtain 
\begin{align}
&\rE_{M_p} P^{\cP_{\vec{n}}}_{ph, min,\leftrightarrow | M_p,u}\nonumber\\
\le &
\sum_{t=0}^{J^1}
p_u(\frac{t}{J^1})
\min\left\{2^{J^1 \overline{h}(\frac{t}{J^1})+J^0+J^2+J^4+J^5-m},
1\right\}
.\label{2-23-2}
\end{align}
Thus, 
from (\ref{2-23-1}) and (\ref{2-23-2}),
the average error probability 
$P^{\cP_{\vec{n}}}_{ph, min,\leftrightarrow| M_p}$
satisfies that
\begin{align}
&\rE_{M_p} P^{\cP_{\vec{n}}}_{ph, min,\leftrightarrow | M_p}\nonumber\\
\le &
\sum_{t=0}^{J^1}
p(\frac{t}{J^1})
\min\left\{2^{J^1 \overline{h}(\frac{t}{J^1})+J^0+J^2+J^4+J^5-m},
1\right\}.
\label{2-23-3}
\end{align}
Thus, we can derive the same argument as subsection \ref{s3d}.
Here, the choice of the sacrifice bit size 
$m$ depends only on the data $\cD_i$ and $\cD_e$.
If we choose the sacrifice bit size $m$ using information $u$,
there is a possibility to improve the above evaluation.

\section{Asymptotic key generation rate}\label{s4}
\subsection{Asymptotic key generation rate with dark count effect}
From the discussion of the precious section,
if we choose the number of sacrifice bits $m$ as a larger number than
$J^1 \overline{h}(r^1)+J^2+J^4+J^5$ in the forward case,
our final key is asymptotically secure.
Hence, we call
$J^1 \overline{h}(r^1)+J^2+J^4+J^5
=N- J^1(1-\overline{h}(r^1))-(J^0+J^3)$ 
the initial Eve's information in the forward case.
Also, 
$J^1 \overline{h}(r^1)+J^0+J^2
=N- J^1(1-\overline{h}(r^1))-(J^3+J^4+J^5)$ is called
the initial Eve's information in the reverse case.
Thus, 
the asymptotic key generation (AKG) rates 
for the detected pulses
of the forward and reverse cases
are equal to 
\begin{align}
&\frac{J^1(1- \overline{h}(r^1))+J^0+J^3}{N}- 
(1-\eta(s_{\nu,+}))
\label{11-9-1}\\
&\frac{J^1(1- \overline{h}(r^1))+J^3+J^4+J^5}{N}- 
(1-\eta(s_{\nu,+}))
\label{11-9-2},
\end{align}
respectively,
when 
$\eta(s_{\nu,+})$ is 
the coding rate of the classical error correction code,
where
$N:=\sum_{i=0}^5 J_i$
and 
$s_{\nu,+}$ is the average error probability of the detected pulses.

In the asymptotic case,
$\frac{J^3+J^4+J^5}{N}$ 
and $\frac{J^0+J^3}{N}$ 
converge to $p_D$ and $\nu(0)p_0$ in probability, respectively,
where $p_0$ is the counting rate of the vacuum pulse
and 
$p_D$ is the rate of the dark counts among sent pulses.
Thus, when our pulse is generated by the distribution $\nu$,
the initial Eve's informations in the forward and reverse cases
are equal to 
\begin{align}
&
N 
(1-\frac{\nu(1) q^1 (1-\overline{h}(r^1))}{p_{\nu,+}}
-\frac{\nu(0) p_0}{p_{\nu,+}})\label{22-11}
\\
&
N (1-\frac{ \nu(1) q^1 (1-\overline{h}(r^1) )}{p_{\nu,+}}
-\frac{p_D}{p_{\nu,+}}),\label{22-10}
\end{align}
respectively,
where $p_{\nu,+}$ is the counting rate of 
the pulse with the $+$ basis generated by the distribution $\nu$,
$q^1$ is the counting rate of the single-photon states
except for dark counts,
and
$r^1$ is the error rate of the $\times$ basis
among the single-photon states
detected except for dark counts.
Hence, 
two important rates $q^1$ and $r^1$ are needed to be estimated.

By taking into account the counting rate $p_{\nu,+}$,
the AKG rates 
for the sent pulses of the forward and reverse cases
are equal to 
\begin{align}
I_{\to}:=
&\frac{
\nu(1) q^1 (1-\overline{h}(r^1))
+\nu(0) p_0- p_{\nu,+}(1-\eta(s_{\nu,+}))}{2}\\
I_{\leftarrow}:=
&\frac{
\nu(1) q^1 (1-\overline{h}(r^1))
+p_D- p_{\nu,+}(1-\eta(s_{\nu,+}))}{2},
\end{align}
respectively,
where $s_{\nu,+}$ 
is the error rate of pulses 
generated with the distribution $\nu$ in the $+$ basis.
These rates are equal to those conjectured by BBL\cite{BBL}.
Hence, the difference between $\frac{\nu(0) p_0}{2}$ and $\frac{p_D}{2}$ 
gives those of the forward and reverse cases.

By applying GLLP\cite{GLLP}-ILM\cite{ILM} formulas, 
the AKG rate is equal to 
\begin{align}
I_{GLLP-ILM}:=
\frac{1}{2}\left(\nu(1) \overline{q^1} (1-\overline{h}(\overline{r^1}))
- p_{\nu,+}(1-\eta(s_{\nu,+}))
\right)\nonumber,
\end{align}
where
$\overline{q^1}$ 
is the rate of all detected single-photon states 
(containing states detected by dark counts),
and $\overline{r^1}$ is the error rate 
among all detected single-photon states in the $\times$ basis \cite{LMC}.
These are calculated as
\begin{align*}
\overline{q^1}&= q^1 +p_D \\
\overline{r^1}&= \frac{r^1 q^1 +\frac{1}{2} p_D}{q^1 +p_D}.
\end{align*}
If we do not take into account the effect of dark counts,
the AKG rates 
of the forward and reverse cases are calculated to 
\begin{align}
\overline{I}_{\to}:=
&\frac{
\nu(1) \overline{q^1} (1-\overline{h}(\overline{r^1}))
+\nu(0) p_0- p_{\nu,+}(1-\eta(s_{\nu,+}))
}{2}\nonumber\\
\overline{I}_{\leftarrow}:=
&\frac{
\nu(1) \overline{q^1} (1-\overline{h}(\overline{r^1}))
- p_{\nu,+}(1-\eta(s_{\nu,+}))
}{2}
\nonumber,
\end{align}
respectively.
The AKG rate $\overline{I}_{\to}$
was 
conjectured by Lo \cite{Lo}, and 
proved by Koashi \cite{Koashi} independently.

The discussion in subsection \ref{s3e} implies
that 
the AKG rate 
\begin{align*}
I_{\leftrightarrow}:=\frac{
\nu(1) q^1 (1-\overline{h}(r^1))
+\nu(0)p_D
- p_{\nu,+}(1-\eta(s_{\nu,+}))
}{2}
\end{align*}
can be attained by two-way error correction\footnote{There is a 
possibility that to improve this bound if our two-way error correction code 
is chosen carefully.}.
Assuming that the coding rate of two-way error correction is
equal to that of one-way error correction,
we compare these AKG rates.
Since
$
\nu(1) q^1 (1-\overline{h}(r^1))+\nu(0)p_D
\ge \nu(1) q^1 (1-\overline{h}(r^1))
= \nu(1) (q^1 (1-\overline{h}(r^1)) +p_D (1-
\overline{h}(\frac{1}{2})))
\ge \nu(1) (q^1+p_D) (1-\overline{h}(
\frac{r^1 q^1 +\frac{1}{2} p_D}{q^1 +p_D}))
= \nu(1) \overline{q^1} (1-\overline{h}(\overline{r^1}))$, we have
\begin{align*}
I_{\to} &\ge \overline{I}_{\to} \ge I_{GLLP-ILM} \\
I_{\leftarrow} &\ge I_{\leftrightarrow} \ge \overline{I}_{\leftarrow} \ge I_{GLLP-ILM}\\
I_{\to} & \ge I_{\leftrightarrow}.
\end{align*}

\subsection{Mixture of the vacuum state and the single-photon state}
First, we assume that $p_S=0$.
Now, we consider 
the distribution $\nu$ 
taking probabilities only in the vacuum state and the single-photon state.
Then, $q^1$ and 
the error rate $r^1=r^1_{\times}$ of the $\times$ basis
can be solved from 
the counting  rate $p_0$ of the vacuum states,
the counting  rate $p_{\nu,\times}$ of the pulses 
generated by $\nu$ in the $\times$ basis,
and the error rate $s_{\nu,\times}$ of the same pulses as follows.
Since 
$q^1$ and $r^1_{\times}$ satisfy the equations:
\begin{align*}
p_{\nu,\times}
&= \nu(0)p_0+\nu(1)(p_D+q^1) \\
s_{\nu,\times}p_{\nu,\times}
&= \frac{1}{2}\nu(0)p_0+\nu(1)(\frac{1}{2}p_D + r^1_{\times} q^1 ) ,
\end{align*}
we obtain 
\begin{align*}
q^1 &=
\frac{p_{\nu,\times}-\nu(0)p_0}{\nu(1)}
-p_D\\
r^1_{\times} &=
\frac{s_{\nu,\times}p_{\nu,\times}-\frac{1}{2}\nu(0)p_0
-\frac{1}{2}\nu(1)p_D }
{p_{\nu,\times}-\nu(0)p_0-\nu(1)p_D}.
\end{align*}
Note that
the counting  rate $p_{\nu,+}$ of the pulses 
generated by $\nu$ in the $+$ basis
coincides with 
the counting  rate $p_{\nu,\times}$ of the pulses 
generated by $\nu$ in the $\times$ basis.
In the case of $p_S\neq 0$,
$r^1_{\times}$ can be calculated as
\begin{align*}
r^1_{\times} =
\frac{\frac{s_{\nu,\times}p_{\nu,\times}-\frac{1}{2}\nu(0)p_0
-\frac{1}{2}\nu(1)p_D }
{p_{\nu,\times}-\nu(0)p_0-\nu(1)p_D}-p_S}{1-2p_S}.
\end{align*}
This is because 
when ${r^1}'$ is
the error probability among the single-photon states detected 
except for dark counts,
the relation ${r^1}'= p_S (1-r^1)+(1-p_S)r^1$ holds.

\subsection{Approximate single-photon state}
Now, 
in the case of $p_S=0$,
we discuss
the distribution $\nu$ 
taking probabilities not only in the vacuum state and the single-photon state
but also in multi-photon states.
Approximate single-photon state has this form.
When we can generate pulses only with the distribution $\nu$,
we have to 
treat 
the rates 
$q^2_{\times}$ and $q^2_{+}$ 
of counuts except for dark counts
of the multi-photon states 
in the $\times$ and $+$ bases
and 
the error rates 
$r^2_{\times}$ and $r^2_{+}$ 
of the multi-photon states detected 
in the $\times$ and $+$ bases
except for dark counts
as unknown parameters 
as well as 
the rate $q^1$ of counts except for dark counts
of the single-photon states 
and 
the error rates 
$r^1_{\times}$ and $r^1_{+}$ 
of the $\times$ basis and the $+$ basis
of the single-photon states detected except for dark counts.
Thus, the following equations hold:
\begin{align}
&p_{\nu,\times} \nonumber\\
=& \nu(0)p_0+\nu(1)(p_D+q^1) +\nu(2)(p_D+q^2_{\times}) \label{1-29-11}\\
&p_{\nu,+}\nonumber\\
=& \nu(0)p_0+\nu(1)(p_D+q^1) +\nu(2)(p_D+q^2_+) \label{1-29-12}\\
&s_{\nu,\times}p_{\nu,\times}\nonumber\\
=& \frac{1}{2}\nu(0)p_0+\nu(1)(\frac{1}{2}p_D + r^1_{\times} q^1 ) 
+\nu(2)(\frac{1}{2}p_D + r^2_{\times} q^2_{\times} ) \label{1-29-13}\\
&s_{\nu,+}p_{\nu,+}\nonumber\\
=& \frac{1}{2}\nu(0)p_0+\nu(1)(\frac{1}{2}p_D + r^1_{+} q^1 ) 
+\nu(2)(\frac{1}{2}p_D + r^2_{+} q^2_{+} ) \label{1-29-14},
\end{align}
where $q^1,q^2_{\times}$, and $q^2_{+}$ 
belong to the interval $[0,1-p_D]$,
and 
$r^1_{\times}$, $r^1_{+}$, $r^2_{\times}$, and $r^2_{+}$ 
do to the interval $[0,1]$.
The AKG rate is characterized by 
the minimum value of 
$q^1 (1-\overline{h}(r^1_{\times}))$
with these conditions.
Since it is difficult to calculate this minimum,
we treat the symmetric case, i.e., the case:
\begin{align}
p_{\nu,\times}=p_{\nu,+},\quad
s_{\nu,\times}=s_{\nu,+}\label{1-29-5}.
\end{align}
Then, the minimum $q^1_{\min}$ of $q^1$ 
and the maximum $r^1_{\max}$ of $r^1_{\times}$ are given as
\begin{align*}
q^1_{\min}&=
\frac{p_{\nu,\times}-p_0\nu(0)-\nu(2)}{\nu(1)}-p_D\\
r^1_{\max}&=
\frac{s_{\nu,\times}p_{\nu,\times}
-\frac{1}{2}p_0\nu(0)
-\frac{1}{2}p_D\nu(1)
-\frac{1}{2}p_D\nu(2)
}
{p_{\nu,\times}-p_0\nu(0)
- p_D\nu(1)-\nu(2)}.
\end{align*}
The minimum $q^1_{\min}$ and 
and the maximum $r^1_{\max}$ 
are realized simultaneously
when $q^2_{\times}=1-p_D, 
r^2_{\times}=0$.
The minimum of 
$q^1 (1-\overline{h}(r^1_{\times}))$
is equal to 
$q^1_{\min} (1-\overline{h}(r^1_{\max}))$.

Next, we consider how much 
AKG rate can be improved when
we send pulses generated by different distributions.
For this purpose, 
we focus on the maximum 
$q^1_{\max}$ of $q^1$ 
and the minimum $r^1_{\max}$ of $r^1_{\times}$,
which are calculated as
\begin{widetext}
\begin{align*}
q^1_{\max}&=
\frac{p_{\nu,\times}-p_0\nu(0)-\nu(2)p_D}{\nu(1)}-p_D\\
r^1_{\min}&\le \tilde{r}^1_{\min}:=
\frac{s_{\nu,\times}p_{\nu,\times}
-\frac{1}{2}p_0\nu(0)
-\frac{1}{2}p_D\nu(1)
-\frac{1}{2}p_D\nu(2)
-(1-p_D)\nu(2)
}
{p_{\nu,\times}-p_0\nu(0)- p_D\nu(1)-\nu(2)p_D}.
\end{align*}
The difference between the maximum and the minimum are 
given as
\begin{align*}
&q^1_{\max}-q^1_{\min}
=
\frac{\nu(2)(1-p_D)}{\nu(1)}\\
& r^1_{\max} -r^1_{\min}
\le
r^1_{\max} -\tilde{r}^1_{\min} \\
\le &
r^1_{\max} -
\frac{s_{\nu,\times}p_{\nu,\times}
-\frac{1}{2}p_0\nu(0)
-\frac{1}{2}p_D\nu(1)
-\frac{1}{2}p_D\nu(2)
-(1-p_D)\nu(2)
}
{p_{\nu,\times}-p_0\nu(0)- p_D\nu(1)-\nu(2)}
(
1-
\frac{\nu(2)(1-p_D)}
{p_{\nu,\times}-p_0\nu(0)- p_D\nu(1)-\nu(2)}
)\\
=&
\frac{(1-p_D)\nu(2)}
{p_{\nu,\times}-p_0\nu(0)- p_D\nu(1)-\nu(2)}
\left(1+
\frac{
s_{\nu,\times}p_{\nu,\times}
-\frac{1}{2}p_0\nu(0)
-\frac{1}{2}p_D\nu(1)
-\frac{1}{2}p_D\nu(2)
-(1-p_D)\nu(2)
}
{p_{\nu,\times}-p_0\nu(0)- p_D\nu(1)-\nu(2)}\right),
\end{align*}
\end{widetext}
where 
the inequality 
$\frac{a}{b+x}
=
\frac{a}{b}
\frac{1}{1+\frac{x}{b}}
\ge
\frac{a}{b}
(1-\frac{x}{b})$ is applied in the case of
$a= s_{\nu,\times}p_{\nu,\times}
-\frac{1}{2}p_0\nu(0)-\frac{1}{2}p_D\nu(1)-\frac{1}{2}p_D\nu(2)-(1-p_D)\nu(2)$,
$b=p_{\nu,\times}-p_0\nu(0)- p_D\nu(1)-\nu(2)$, and
$x=\nu(2)(1-p_D)$.
Hence, when these differences are small relatively with 
$q_{\min}$ and $r_{\max}$,
the AKG rate cannot be improved so much 
even though we send pulses generated by different distributions.
For example, 
$(1-p_D)\nu(2)$ is small enough when 
the generated pulse is close enough to the single-photon.

When the symmetric assumption (\ref{1-29-5}) does not hold,
the conditions (\ref{1-29-12}) and (\ref{1-29-14}) are added with
the conditions (\ref{1-29-11}) and (\ref{1-29-13}).
The maximums $q^1_{\max}$ and $r^1_{\max}$ become small,
and the minimums $q^1_{\min}$ and $r^1_{\min}$ become large.
Hence, the following relations also hold even in the non-symmetric case:
\begin{widetext}
\begin{align*}
q^1_{\max}-q^1_{\min}
\le &
\frac{\nu(2)(1-p_D)}{\nu(1)}\\
r^1_{\max} -r^1_{\min}
\le &
\frac{(1-p_D)\nu(2)}
{p_{\nu,\times}-p_0\nu(0)- p_D\nu(1)-\nu(2)}
\left(1+
\frac{
s_{\nu,\times}p_{\nu,\times}
-\frac{1}{2}p_0\nu(0)
-\frac{1}{2}p_D\nu(1)
-\frac{1}{2}p_D\nu(2)
-(1-p_D)\nu(2)
}
{p_{\nu,\times}-p_0\nu(0)- p_D\nu(1)-\nu(2)}\right).
\end{align*}
\end{widetext}

\section{Detail analysis on Eve's attack}\label{s5}
\subsection{Reduction to three-dimensional outcome channel}
We prove that any Eve's attack can be reduced by the 
attack discussed in Section \ref{s3}.
Of course, in the following discussion contains the case when
the frame of Alice does not coincide with that of Bob.
Since Alice sends $N$ pulses and Bob receives $N$ pulses,
Eve's operation can be described by
a CP-TP map $\cE_N$ from the system 
$\cH^{\otimes N}$ to the system $\cH^{\otimes N}$.
This description contains the loss of the communication channel.
Even if the detector has the loss,
if the loss does not depend on the measurement basis,
the security is guaranteed by our discussion.

In order to reduce the output system $\cH$
to three-dimensional system $\cH_0 \oplus \cH_1 $,
we modify our protocol as follows:
In the measurement with the $+$ basis,
Bob performs the measurement 
$\{|n,m\rangle\}_{n,m}$.
When $(0,0)$ is measured, he decides his final outcome to be $\emptyset$.
When $(n,m)$ is measured, he does his final outcome to be 
$0$ with the probability $\frac{n}{n+m}$, 
and $1$ with the probability $\frac{m}{n+m}$.
This POVM with three outcomes is denoted by 
$\{\tilde{M}_{\emptyset},\tilde{M}_0,\tilde{M}_1\}$
on the system $\cH$.
First, we discuss the security based on the POVM
$\{\tilde{M}_{\emptyset},\tilde{M}_0,\tilde{M}_1\}$,
and after this discussion, we treat the security with 
the POVM $\{M_{\emptyset},M_0,M_1\}$,
which is given in Section \ref{s2}.

The stochastic behavior of the outcome of 
the POVM $\{\tilde{M}_{\emptyset},\tilde{M}_0,\tilde{M}_1\}$ 
is described by the POVM
$\{
M_{\emptyset}':=|0\rangle \langle 0|,
M_{0}':=|1,0\rangle \langle 1,0|,
M_{1}':=|0,1\rangle \langle 0,1|\}$
on the system $\cH_0\oplus \cH_1$ 
and the TP-CP map $\cE$,
which are defined as
\begin{align*}
\Tr \tilde{M}_i \rho&=\Tr e_i' \cE(\rho)\\
\cE(\rho)&:=
P_{\cH_0}\rho P_{\cH_0}
+P_{\cH_1}\rho P_{\cH_1}
+\sum_{n=2}^{\infty} \cE^n(P_{\cH_n}\rho P_{\cH_n}),
\end{align*}
where
the TP-CP map $\cE^n$ from the 
$n$-photon system $\cH_n$ 
(which is equal to the $n$-th symmetric space)
to the system $\cH_1$ 
is defined 
by embedding the state $\rho$ on the $n$-th symmetric space $\cH_n$ into 
$n$-tensor product system $\cH_1^{\otimes n}$ as
\begin{align*}
\cE^n(\rho):= \Tr_{2, \ldots, n} \rho,
\end{align*}
where $\Tr_{2, \ldots, n}$ means taking partial trace 
concerning $2$-th - $n$-th subsystems.

Hence, by extending the system under Eve's control,
the environment of the TP-CP map $\cE$ 
can be regarded to be under Eve's control.
The channel from Alice to Bob
can be described 
as the TP-CP map $\cE^{\otimes N}\circ \cE_N$ from the system 
system $\cH^{\otimes N}$ to the system $(\cH_0\oplus \cH_1)^{\otimes N}$.
We also assume a stronger ability of Eve, i.e.,
all states 
$|n,0\rangle$, $|0,n\rangle$ $|n,0,\times\rangle$, $|0,n,\times\rangle$ can
be distinguished by Eve for $n \ge 2$.
Then, each pulse can be described as a state on the system 
$\cH':= \cH_0 \oplus \cH_1 
\oplus (\oplus_{n=2}^{\infty} \cH_{n,+})
\oplus (\oplus_{n=2}^{\infty} \cH_{n,\times})$,
where
the space $\cH_{n,+}$ is spanned by
$\{|n,0\rangle, |0,n\rangle\}$,
and the space $\cH_{n,\times}$ 
is spanned by
$\{|n,0,\times\rangle, |0,n,\times\rangle\}$.
Then, 
the channel $\cE^{\otimes N}\circ \cE_N$ from Alice to Bob
can be regarded as 
a TP-CP map from the system ${\cH'}^{\otimes N}$ to the
system $(\cH_0\oplus \cH_1)^{\otimes N}$,
which is denoted by $\cE_N'$ in the following.

Now, we focus on following two pinching maps:
\begin{align*}
\cE_a (\rho):= &
P_{\cH_1}\rho P_{\cH_1}+
|0,0\rangle \langle 0,0| \rho |0,0\rangle \langle 0,0|\\
&+
\sum_{n=2}^{\infty}
|n,0\rangle \langle n,0| \rho |n,0\rangle \langle n,0|\\
&\qquad
+|0,n\rangle \langle 0,n| \rho |0,n\rangle \langle 0,n|\\
&\qquad
+|n,0,\times\rangle \langle n,0,\times| \rho 
|n,0,\times\rangle \langle n,0,\times| \\
&\qquad
+|0,n,\times\rangle \langle 0,n,\times| \rho 
|0,n,\times\rangle \langle 0,n,\times|,\\
\cE_b (\rho):= &P_{\cH_0}\rho P_{\cH_0}+P_{\cH_1}\rho P_{\cH_1}.
\end{align*}
By extending the system controlled by Eve,
the channel from Alice to Bob can be regarded as
a TP-CP map $\cE_N:= \cE_b\circ \cE_N' \circ \cE_a$ from 
the system ${\cH'}^{\otimes N}$ to the 
system $(\cH_0\oplus \cH_1)^{\otimes N}$,
due to the forms of the measurement by Bob and the states sent by Alice.
\subsection{Discrete twirling}
In order to define discrete twirling, 
we define the operators $\sX$ and $\sZ$ on the system 
system $\cH'$ by
\begin{align*}
\sX |0\rangle &= |0\rangle \\
\sX |\uparrow\rangle &= |\downarrow\rangle , \quad
\sX |\downarrow\rangle = |\uparrow\rangle \\
\sX |n,0\rangle & = |0,n\rangle , \quad
\sX |0,n\rangle = |n,0\rangle \\
\sX |n,0,\times\rangle &= |n,0,\times\rangle, \quad
\sX |0,n,\times\rangle = - |0,n,\times\rangle \\
\sZ |0\rangle &= |0\rangle \\
\sZ |\uparrow\rangle &= |\downarrow\rangle , \quad
\sZ |\downarrow\rangle = (-1)|\uparrow\rangle \\
\sZ |n,0\rangle &= |n,0\rangle , \quad
\sZ |0,n\rangle = -|0,n\rangle \\
\sZ |n,0,\times\rangle & = |0,n,\times\rangle, \quad
\sZ |0,n,\times\rangle = |n,0,\times\rangle ,
\end{align*}
and the operators $\sX^x$ and $\sZ^z$ for 
$x,z \in \bF_2^N$ by
\begin{align*}
\sX^{x}& = \sX^{x_1} \otimes \cdots\otimes 
\sX^{x_N} , \quad
\sZ^{z}= Z^{z_1} \otimes \cdots\otimes Z^{z_N} .
\end{align*}
It is known that if and only if the relation 
\begin{align}
(\sX^x \sZ^z )^{\dagger}
\cE_N(\sX^x \sZ^z \rho (\sX^x \sZ^z )^{\dagger})
\sX^x \sZ^z 
=\cE_N(\rho)\label{1-19-1}
\end{align}
holds for any $x,z \in \bF_2^N$,
$\cE_N$ has the form of (\ref{1-19-2}).
Now, we define 
the discrete twirling $\cE_N$ 
of the map $\cE_N$:
\begin{align}
\cE_N(\rho):=
\frac{1}{2^{2N}}
\sum_{x,z \in \bF_2^N}
(\sX^x \sZ^z )^{\dagger}
\cE_N(\sX^x \sZ^z \rho (\sX^x \sZ^z )^{\dagger})
\sX^x \sZ^z .\label{1-19-5}
\end{align}
The operation of `discrete twirling'
corresponds to the following operation:
First, Alice generates
two random numbers $x,z \in \bF_2^N$,
performs the operation $\sX^x \sZ^z$,
and sends the state via the channel $\cE_N$,
and the classical information 
$x,z \in \bF_2^N$ via the public channel.
Next, Bob performs the inverse operation 
$(\sX^x \sZ^z )^{\dagger}$ to the received system.
Since the TP-CP map $\cE$ has the covariance (\ref{1-19-1}),
$\cE$ has the form (\ref{1-19-2}).
However, when Eve's system is extended,
the implemented channel 
$\cE$ is not ${\cE}_N$ but only ${\cE'}_N$.

\subsection{Forward case}
In the forward error correction case, 
by taking into account the error correction operation,
the raw keys can be regarded to be transmitted via 
the discrete twirling of the original channel.
The operation concerning $x$ in (\ref{1-19-5}) is 
essentially realized by the following operation.
After quantum communication,
Alice generates the random number $X$ 
and sends the classical information 
$Y=M_e Z+X$. Bob regards $X'-Y$ as the final raw key.
On the other hand, the input and output data
are not changed when the operation corresponding to $z$.
If Alice and Bob perform the operation concerning $z$ in (\ref{1-19-5}),
Eve's performance is increased only.
Hence, in the forward case,
the security can be evaluated by analysis on the channel (\ref{1-19-2}).
Note that
since our error is symmetric in the form (\ref{1-19-2}),
the error probability can be estimated from
the random variable $\cD_e$.

When we take into account dark counts,
any channel concerning pulses detected expect for dark counts,
can be described by (\ref{1-19-2}) in the forward case.
Eve can obtain all information when 
the sent pulse is not in the vacuum state but is detected by dark counts
while he cannot obtain any information when 
the sent pulse is in the vacuum state and is detected by dark count.
Hence, the analysis in Section \ref{s3} is valid in the forward case.

\subsection{Reverse case}
We proceed to the case where all sent states are in single-photon 
and the reverse error correction is applied.
Our protocol is given as follows.
First, Alice sends
the information bits $X$ with the $+$ basis to Bob via the quantum channel
$\Lambda$, and Bob obtains the bits $X'$ by measuring the received states
with the $+$ basis.
Bob generates another random number $Z$, and 
sends the classical information $Y=Z+X'$ to Alice,
and Alice regards $X-Y$ as the final raw keys.
Now, we consider the following modified protocol:
Alice generates an entangled pair and sends one part to Bob.
Bob generates another quantum state with the $+$ bit basis,
and performs 
the Bell measurement 
$\{M_{(0,0)},M_{(0,1)},M_{(1,0)},M_{(1,1)}\}$
to the joint system of the received system and his original system,
where
$M_{(0,0)},M_{(0,1)},M_{(1,0)},M_{(1,1)}$ is 
the projection corresponding to 
$\frac{1}{\sqrt{2}}(|0\rangle |0\rangle +|1\rangle |1\rangle),
\frac{1}{\sqrt{2}}(|0\rangle |0\rangle -|1\rangle |1\rangle),
\frac{1}{\sqrt{2}}(|0\rangle |1\rangle +|1\rangle |0\rangle),
\frac{1}{\sqrt{2}}(|0\rangle |1\rangle -|1\rangle |0\rangle)
$, respectively.
Then, he sends his measurement value to Alice.
Alice performs the inverse transformation depending on the data
to her system, and measures it with the $+$ basis.
Since 
the map from Bob's input state to the Alice's final state 
satisfies the covariance (\ref{1-19-1}),
this channel is described by 
the Pauli channel $\cE(\Lambda)$ that is the discrete twirling of $\Lambda$.
The latter protocol is essentially 
equivalent with 
the former protocol with the following modification:
Bob sends Alice $(0,0)$ and $(0,1)$ with the probability $\frac{1}{2}$
in the case of $Y=0$,
and sends Alice $(1,0)$ and $(1,1)$ with the probability $\frac{1}{2}$
in the case of $Y=1$.
Hence, the channel from Bob to Alice in the former case 
can be described by the Pauli channel $\cE(\Lambda)$.
Therefore, without loss of generality, 
we can assume that 
the original map from Alice to Bob can be regarded as a Pauli channel.

Now, in order to treat the loss during the transmission,
we modify the latter protocol as follows.
Alice performs the two-valued measurement $\{T_0,T_1\}$ before 
sending one part of the entangled pair,
and sends the state 
$ \sqrt{T_1}\rho \sqrt{T_1}$ only when $1$ is detected.
In this case, 
the map from Bob's input state to the Alice's final state 
can be described by a Pauli channel.
This protocol is essentially equivalent with 
the modification of the above former protocol with the following modification:
Alice performs 
the two-valued measurement $\{T_0,T_1\}$ before 
sending her state,
and sends the state $ \sqrt{T_1}\rho \sqrt{T_1}$ only when $1$ is detected.
This modification is equivalent with the lossy channel case.

Next, we consider the case where
the number of photons in the input state 
is not fixed and no dark count is detected.
In this case, we assume that Eve can know
Bob's measured value
when $n$-photon state ($n \ge 2$) 
or the vacuum state is transmitted.
It is needed only to describe
the behavior of the counting rates and the error rates,
which are estimated from the random variable $\cD_i$.
Thus, we can assume that the channel from Alice to Bob can be described by
(\ref{1-19-2}) without loss of generality.

Finally, we consider the case with dark counts.
Eve cannot obtain any Bob's information for the bits detected by dark counts.
Hence, 
Eve's information concerning this part 
has no relation with 
the channel from Alice to Bob.
Any description of this part is allowed.
Thus, even if the dark counts exist,
the channel from Alice to Bob can be described 
by (\ref{1-19-2}) without loss of generality.

\subsection{Security with the original POVM}
We compare
the case with the measurement 
$\{\tilde{M}_{\emptyset},\tilde{M}_0,\tilde{M}_1\}$ 
and that with the measurement
$\{M_{\emptyset},M_0,M_1\}$.
The systems controlled by Eve in these two cases 
are identical.
The error probability 
based on the measurement 
$\{M_{\emptyset},M_0,M_1\}$
is larger than 
that based on the measurement 
$\{\tilde{M}_{\emptyset},\tilde{M}_0,\tilde{M}_1\}$.
Thus, the estimate of the error rate $r^1$ with the measurement
$\{M_{\emptyset},M_0,M_1\}$
is larger than
that 
with the measurement
$\{\tilde{M}_{\emptyset},\tilde{M}_0,\tilde{M}_1\}$.
Since a larger estimate of $r^1$ gives a larger size $m$ of sacrifice bits,
the security based on the measurement 
$\{\tilde{M}_{\emptyset},\tilde{M}_0,\tilde{M}_1\}$ 
implies 
the security based on the original measurement 
$\{M_{\emptyset},M_0,M_1\}$.

\section{Concluding remarks}
Applying the relation between Eve's information and phase error
probability, we have derived useful upper bounds 
of eavesdropper's performances, i.e.,
eavesdropper's information and the trace
norm between the Eve's states corresponding to final keys
for the protocol given in section \ref{s2}. 
Here, we have used powerful relations between
the eavesdropper's performances and
the phase error probability.
We have also treated our channel as a TP-CP map on the two-mode bosonic system,
which is the most general framework. Further, our discussion 
has taken into
account the effect of dark counts, which forbids Eve to access Bob's
bits of the pulses detected by dark count. However, our upper bounds
(\ref{1-29-2}) and (\ref{1-29-1}) contain so many random variables
that its detail numerical analysis in the case of phase-randomized
coherent light is very complicated and is separately given in Hayashi
et al \cite{H3}. Hence, the future problem for practical QKD is the
numerical calculations of the bounds (\ref{1-29-2}) and
(\ref{1-29-1}). On the other hand, the concrete calculation of the AKG
rate is another future important topic. Also this topics is separately
discussed in Hayashi \cite{H2}.

We have treated the AKG rate more deeply when the generated imperfect
resource is close to the single-photon. Since this resource is
different from the perfect single-photon, we need the decoy method.
We have compared the case where only the vacuum state is sent as a different
pulse with the case where additional pulses are sent as different
pulses.

This paper has treated only the binary case. However, it is easy to extend
to the $p$-nary case, where $p$ is a prime. In this case, we replace
the two-mode bosonic system and $\bF_2$ by the $p$-mode bosonic system
and $\bF_p$.

\section*{Acknowledgments}
The author would like to thank Professor Hiroshi Imai of the
ERATO-SORST, QCI project for support.
He is grateful to Professor Hiroshi Imai, Dr. Akihisa Tomita, 
Dr. Tohya Hiroshima, and Mr. Jun Hasewaga for useful discussions.
He thanks Dr. Francesco Buscemi for his meaningful comments.
He also thanks the referee for his helpful comments.

\appendix
\section{Proof of Theorem \ref{th1}}\label{stoep}
Now, we prove Theorem \ref{th1}.
More precisely, we show the following.
(1) An element $(x,y)^T \in \bF_2^m \oplus \bF_2^{l}$ 
belongs the image of $(\bX,I)^T$
with the probability $2^{-m}$ if $x \neq 0$ and $y \neq 0$.
(2) it does not belong to the image of the transpose of any 
$l \times (l+m)$ Toeplitz matrices $(\bX,I)$
if $x \neq 0$ and $y = 0$.

Indeed, since (2) is trivial, we will show (1).
For $y=(y_1, \ldots, y_m)$, 
we let $i$ be the minimum index $i$ such that $y_i \neq 0$.
An element $(x,y)^T \in \bF_2^m \oplus \bF_2^{l}$ belongs to 
the image of $(\bX,I)^T$ if and only if $(x,y)= (\bX^T y,y)$,
which written as 
the following $m$ conditions.
\begin{align*}
Y_i y_i &= x_1 - \sum_{j=i+1}^{l} Y_{j} y_j -x_1 \\
Y_{i+1} y_i &= x_2 - \sum_{j=i+1}^{l} Y_{j+1} y_j -x_1 \\
& \vdots \\
Y_{i+m-2} y_i &= x_{m-1} - \sum_{j=i+1}^{l} Y_{j+m-2} y_j -x_1 \\
Y_{i+m-1} y_i &= x_m - \sum_{j=i+1}^{l} Y_{j+m-1} y_j -x_1 
\end{align*}
Now, the random variables $Y_{i+m},\ldots, Y_{l+m-1}$ are fixed.
The $m$-th condition does not depend on the
variables $Y_1, \ldots Y_{i+m-2}$.
Hence, the $m$-th condition only depends on the variable $Y_{i+m-1}$.
Therefore, the $m$-th condition holds with the probability $1/2$.
Similarly, we can show that
the $m-1$-th condition holds with the probability
$1/2$ when the $m$-th condition holds.
Thus, the $l$-th condition and $l-1$-th condition 
hold with $1/2^2$.
Repeating this discussion inductively,
we can conclude that
the all $m$ conditions hold with the probability $2^{-m}$.

\begin{widetext}
\section{Eve's states}\label{a2}
For any Pauli channel $\cE(\rho)=
\sum_{x,z\in \bF_2^l} 
P(x,z) X^x Z^z \rho (X^x Z^z )^{\dagger}$,
the channel to Eve is given by $\cE_E(\rho)$:
\begin{align*}
\cE_E(\rho):=
\sum_{(x,z),(x',z')\in \bF_2^{2l}}
\sqrt{P(x,z)}\sqrt{P(x',z')}
\Tr X^x Z^z \rho (X^{x'} Z^{z'} )^{\dagger}
|(x,z)\rangle \langle(x',z')|.
\end{align*}
If the input state is given by $|y\rangle$ in the $+$ basis,
Eve's state can be evaluated as follows.
\begin{align}
&\cE_E(|y\rangle \langle y|) \nonumber\\
=&
\sum_{
x \in \bF_2^l,z,z'\in \bF_2^{l}}
\sqrt{P(x,z)}\sqrt{P(x,z')}
\langle y| Z^{z-z'} |y\rangle 
|(x,z)\rangle \langle(x,z')| \nonumber\\
=&
\sum_{x \in  \bF_2^l,z,z'\in \bF_2^{l}}
P(x)\sqrt{P(z|x)}\sqrt{P(z'|x)}
(-1)^{(z-z')\cdot y}
|(x,z)\rangle \langle(x,z')|\nonumber\\
=&
\sum_{x \in \bF_2^l}
P(x)|P,y,x\rangle \langle P,y,x|,\label{2-22-1}
\end{align}
\end{widetext}
where we define the vector 
$|P,y,x\rangle $ as
\begin{align*}
|P,y,x\rangle:=
\sum_{z\in \bF_2^{l}}
(-1)^{z\cdot y}\sqrt{P(z|x)}
|(x,z)\rangle .
\end{align*}
That is, Eve's state is the stochastic mixture
of the state 
$|P,y,x\rangle \langle P,y,x|$ with the probability $P(x)$.
Then, Eve loses no information even if she measure the information $x$
concerning the error of the $+$ basis.

\section{Proof of Theorem \ref{t-6-1}}\label{as3}
The inequalities (\ref{1-19-6-2}) and (\ref{1-19-7-2}) are 
proved by Hayashi \cite{hayashi}.
First, we prove the inequalities 
(\ref{1-19-8}), (\ref{1-19-9}),
(\ref{1-19-8-2}), and (\ref{1-19-9-2}).
As is similar to the inequalities 
(\ref{1-19-6-2}) and (\ref{1-19-7-2}),
it is sufficient to show these inequalities
for the corrected channel.
This is because the code $\im M_e/M_e (\Ker M_p)$ with the $+$ basis 
is equivalent with 
the code $(M_e (\Ker M_p))^{\perp}/ (\im M_e)^{\perp}$  with
the $\times$ basis \cite{hayashi}.
Thus, it is sufficient to show
\begin{align}
\min_{y,y' \in \bF_2^l}
F(\cE_E(|y\rangle \langle y|) ,\cE_E(|y'\rangle \langle y'|) )
&\ge 1- 2 P_{ph} \label{1-19-3}
\\
\max_{y,y' \in \bF_2^l}
\|\cE_E(|y\rangle \langle y|) -\cE_E(|y'\rangle \langle y'|) \|_1
&\ge 4 P_{ph} 
\label{1-19-4}\\
\min_{y \in \bF_2^l}
F(\cE_E(|y\rangle \langle y|) ,\rho_{\mix} )
&\ge 1- P_{ph} \label{1-19-3-2}
\\
\max_{y \in \bF_2^l}
\|\cE_E(|y\rangle \langle y|) -\rho_{\mix} \|_1
&\ge  P_{ph} ,
\label{1-19-4-2}
\end{align}
where 
$\rho_{\mix}$ is the maximally mixed state
and the phase error probability $P_{ph}$ is defined as
\begin{align*}
P_{ph}:= \sum_{x,z \in \bF_2^l,z\neq 0}
P(x,z)=\sum_{x \in \bF_2^l}P(x,0)-1.
\end{align*}

Since $\|\rho- \rho'\|_1
\ge 2 (1-F (\rho,\rho'))$,
the inequalities (\ref{1-19-4}) and (\ref{1-19-4-2}) 
follow from (\ref{1-19-3}) and (\ref{1-19-3-2}).
Now, we will prove (\ref{1-19-3}) and (\ref{1-19-3-2}).
Remember the relation (\ref{2-22-1}).
Since
\begin{align*}
& |\langle P,y,x|P,y',x\rangle |
=
|\sum_{z\in \bF_2^{l}}
(-1)^{z\cdot (y'-y)}
P(z|x)| \\
\ge &
P(0|x)
-|\sum_{z\in \bF_2^{l}, z \neq 0}
(-1)^{z\cdot (y'-y)}
P(z|x)| \\
= &
P(0|x)-(1-P(0|x))
=
2 P(0|x)-1,
\end{align*}
the fidelity 
$F(\cE_E(|y\rangle \langle y|) ,\cE_E(|y'\rangle \langle y'|) )$
can be evaluated as
\begin{align*}
& F(\cE_E(|y\rangle \langle y|) ,\cE_E(|y'\rangle \langle y'|) )
=
\sum_{x \in \bF_2^l}
P(x)
|\langle P,y,x|P,y',x\rangle | \\
\ge &
\sum_{x \in \bF_2^l}
P(x)
(2 P(0|x)-1)
=
2 \sum_{x \in \bF_2^l}P(x,0)-1
= 1-2 P_{ph},
\end{align*}
which implies (\ref{1-19-3}).
Since
\begin{align*}
& \langle P,y,x|
\bigl(
\frac{1}{2^l}
\sum_{y'\in \bF_2^{l}}
|P,y',x\rangle \langle P,y',x| \bigr)
|P,y,x\rangle \\
=& \langle P,y,x|
\bigl(
\sum_{z\in \bF_2^{l}}
P(z|x)
|(x,z)\rangle \langle(x,z)|\bigr)
|P,y,x\rangle 
\ge
P(0|x)^2,
\end{align*}
we have
\begin{align*}
& F\Bigl(\cE_E(|y\rangle \langle y|) ,
\sum_{y'\in \bF_2^{l}}
\frac{1}{2^l}
\cE_E(|y'\rangle \langle y'|) \Bigr)\\
=&
\sum_{x \in \bF_2^l}
P(x)
\sqrt{
\langle P,y,x|
\bigl(
\frac{1}{2^l}
\sum_{y'\in \bF_2^{l}}
|P,y',x\rangle \langle P,y',x| \bigr)
|P,y,x\rangle 
}\\
\ge &
\sum_{x \in \bF_2^l}
P(x)
P(0|x)
= 
\sum_{x \in \bF_2^l}P(x,0)
= 1- P_{ph},
\end{align*}
which implies (\ref{1-19-3-2}).

Next, we show (\ref{1-31-1}).
The discrimination on the set of states
$\{
\cE_E(|y\rangle \langle y|) \}_{y\in \bF_2^l}$ can be reduced to
The discrimination on the set of states
$\{|P,y,x\rangle \langle P,y,x| \}_{y\in \bF_2^l}$.
This set has a symmetry concerning the action of 
$y' \in \bF_2^l$ as 
$U_{y'}:=\sum_{z\in \bF_2^{l}}
(-1)^{z \cdot y'}
|(x,z)\rangle \langle(x,z)|$.
Each one-dimensional subspace 
spanned by $|(x,z)\rangle $ is different representation subspace 
in the space spanned by
$\{|(x,z)\rangle \}_{z\in \bF_2^{l}}$.
From Holevo\cite{HoC}'s theory of covariant estimator,
the minimum average error is given by the following covariant POVM
$\{2^{-l}U_{y}|\phi\rangle \langle \phi|U_{y}^\dagger\}$,
where 
$|\phi\rangle=\sum_{z\in \bF_2^{l}}
e^{i \theta_z}|(x,z)\rangle$.
Then, the correct-decision probability 
is given as
\begin{align*}
2^{-l}|\langle P,0,x|\phi\rangle|^2
= 
2^{-l}\Bigl|\sum_{z\in \bF_2^{l}}
\sqrt{P(z|x)}e^{i \theta_z}\Bigr|^2.
\end{align*}
Its maximal value is $(\sum_{z\in \bF_2^{l}}
\sqrt{P(z|x)}\sqrt{2^{-l}}
)^2$,
which is attained when $e^{i \theta_z}=1$.
Therefore,
the optimal correct-decision probability of the set 
$\{
\cE_E(|y\rangle \langle y|) \}_{y\in \bF_2^l}$ 
is equal to 
$\sum_{x \in \bF_2^l}
P(x)(\sum_{z\in \bF_2^{l}}\sqrt{P(z|x)}
\sqrt{2^{-l}})^2$.

Since 
$(\sum_{z\in \bF_2^{l}}\sqrt{P(z)}
\sqrt{2^{-l}})^2$ 
is the fidelity between 
the uniform distribution and the distribution $P$,
the joint concavity of the fidelity guarantees that
the concavity of 
$(\sum_{z\in \bF_2^{l}}\sqrt{P(z)}
\sqrt{2^{-l}})^2$ concerning the distribution $P$.
Thus,
\begin{align*}
& \sum_{x \in \bF_2^l}
P(x)
\Bigl(\sum_{z\in \bF_2^{l}}
\sqrt{P(z|x)}\sqrt{2^{-l}}\Bigr)^2 \\
\le &
\Bigl(\sum_{z\in \bF_2^{l}}
\sqrt{
\sum_{x \in \bF_2^l}P(x,z)}\sqrt{2^{-l}}\Bigr)^2.
\end{align*}
The concavity guarantees that
\begin{align*}
& \max_{P:P(0)=1-P_{ph}}
\bigl(\sum_{z\in \bF_2^{l}}
\sqrt{P(z)}\sqrt{2^{-l}}\bigr)^2\\
=&
\biggl(\sum_{z\in \bF_2^{l}}
\sqrt{1-P_{ph}}\sqrt{2^{-l}}
+
(2^{l}-1)
\sqrt{\frac{P_{ph}}{2^{l}-1}}
\sqrt{2^{-l}}
\biggr)^2\\
=&
\bigl(
\sqrt{P_{ph}}
\sqrt{1-2^{-l}}
+
\sqrt{1-P_{ph}}
\sqrt{2^{-l}}
\bigr)^2,
\end{align*}
which implies (\ref{1-31-1}).

Applying the concavity 
of
$(\sum_{z\in \bF_2^{l}}\sqrt{P(z)}
\sqrt{2^{-l}})^2$ 
between 
the distributions
$(1-P_{ph},\frac{P_{ph}}{2^{l}-1}, \ldots, \frac{P_{ph}}{2^{l}-1})$
and 
$(1-{P'}_{ph},\frac{{P'}_{ph}}{2^{l}-1}, \ldots, \frac{{P'}_{ph}}{2^{l}-1})$,
we obtain the concavity of 
$\bigl(
\sqrt{P_{ph}}
\sqrt{1-2^{-l}}
+
\sqrt{1-P_{ph}}
\sqrt{2^{-l}}
\bigr)^2$ concerning $P_{ph}$.


\section{Proof of Theorem \ref{t-6-2}}\label{as3-2}
Since the condition $Z \in (M_e \Ker M_p)^{\perp}$
is equivalent with the condition
$M_e^{T} Z \in \im M_p^T$ for $Z \in \bF_2^{N}$,
the condition (\ref{2-9-1}) is equivalent with the condition:
\begin{align*}
\rP\{
Z \in (M_e \Ker M_p)^{\perp}
\}\le 2^{-m} \hbox{ for } \forall Z \in \bF_2^{N}\setminus (\im M_e)^{\perp}.
\end{align*}
Hence, Theorem \ref{t-6-2} is essentially equivalent with
the following proposition, which we will prove.
\begin{prop}\label{t-16-1}
Let $C_1 \subset \bF_2^N$ be an $l+m$-dimensional code.
We choose an $m$-dimensional subcode $C_2(X) \subset C_1$
satisfying the following condition:
\begin{align}
\rP_X \{X| x \in C_2(X)^{\perp}\} 
\le 2^{-m}
\hbox{ for }\forall x \in 
\bF_2^n\setminus C_1^{\perp},
\label{11-16-2}
\end{align}
where $X$ is the random variable describing the stochastic behavior.
Alice sends the information $C_2(X)^{\perp}/ C_1^{\perp}$
via the following channel.
Here, when she wants to send the information 
$[x]_{C_1^{\perp}}\in C_2(X)^{\perp}/ C_1^{\perp}$,
she chooses $x'$ among $x + C_1^{\perp}$ with 
the equal probability $2^{l+m-N}$.
The total $N$ bits can be divided to the following three parts:
\begin{description}
\item[$n_0$ bits:] no noise (0th part).
\item[$n_1$ bits:] at most $t$ bits will be changed (1st part).
\item[$n_2$ bits:] no assumption (2nd part).
\end{description}
Using the above classification,
Bob recovers the original information 
for received information $y$ by the following way.
First, he defines the element $\Gamma (y) \in C_2(X)^{\perp}$
by
\begin{align*}
\Gamma (y):=
\argmax_{z\in C_2(X):
\pi_0(y)= \pi_0(z)}|\pi_1(y-z)|,
\end{align*}
where
$\pi_i$ is the projection 
to the above $n_i$ bits.
Next, Bob recovers the information 
$[\Gamma (y)]_{C_1^{\perp}} \in C_2(X)^{\perp}/ C_1^{\perp}$.

Then, we obtain
\begin{align}
& \rE_X
{p_x} \{[\Gamma (y)]_{C_1^{\perp}}\neq [x]_{C_1^{\perp}}\}
\nonumber\\
\le &
2^{n_1 \overline{h}(\frac{t}{n_1})+n_2 -m}
\label{11-16-4}
\end{align}
for any $[x]_{C_1^{\perp}} \in C_2(X)^{\perp}/ C_1^{\perp}$,
where
$p_x$ is the conditional distribution 
describing the distribution of the output $y$
with the input $x$
satisfying the above condition.
\end{prop}
\begin{proof}
From the linearity,
\begin{align*}
{p_x} \{[\Gamma (y)]_{C_1^{\perp}}\neq [x]_{C_1^{\perp}}\}
=&
{p_0} \{[\Gamma (y)]_{C_1^{\perp}}\neq [0]_{C_1^{\perp}}\} \\
=&
{p_0} \{
\Gamma (y)
\notin C_1^{\perp}\}.
\end{align*}
Hence, it is enough to show
\begin{align}
\rE_X
\rP_{p_0} \{
\Gamma (y)
\notin C_1^{\perp}\}
\le
2^{n_1 \overline{h}(\frac{t}{n_1})+n_2 -m}.
\label{11-16-6}
\end{align}
When the original massage is $0$,
the received signal $y$ 
satisfies the 
conditions $\pi_0(y)=0$ and $|\pi_1(y)|\le t$, i.e.,
the distribution $p_0$ has positive probability only on the set 
${\cal Y}:=\{y|\pi_0(y)=0, |\pi_1(y)|\le t\}$.
Then,
\begin{align}
& \rE_X
{p_0} \{
\Gamma (y)
\notin C_1^{\perp}\} \nonumber\\
=&
\rE_X
{p_0} \{y|
\exists z \in C_2(X)^{\perp}\setminus C_1^{\perp}
 \hbox{s.t.}
|\pi_1(z)- \pi_1(y)|\le |\pi_1(y)|\} \nonumber\\
\le &
\rE_X p_0
\{y|\exists z \in C_2(X)^{\perp} \setminus C_1^{\perp}
\hbox{s.t.}
|\pi_1(z)- \pi_1(y)|\le t\}\nonumber\\
= &
\sum_{y\in {\cal Y}}
p_0 (y)
\rP_X 
\left\{X
\left|
\begin{array}{l}
\exists z \in C_2(X)^{\perp} \setminus C_1^{\perp}\\
\hbox{s.t.}
|\pi_1(z)- \pi_1(y)|\le t
\end{array}
\right.\right\}
\nonumber\\
\le &
\sum_{y\in {\cal Y}}p_0 (y)
\sum_{z:|\pi_1(z)- \pi_1(y)|\le t }
\rP_X
\{X|z \in C_2(X)^{\perp} \setminus C_1^{\perp}\} \nonumber\\
\le &
\sum_{y\in {\cal Y}}p_0 (y)
\sum_{z:|\pi_1(z)- \pi_1(y)|\le t }2^{-m}
\label{4-25-1}\\
\le &
\sum_{y\in {\cal Y}}p_0 (y)2^{n_1 \overline{h}(\frac{t}{n_1})+n_2 -m}
=2^{n_1 \overline{h}(\frac{t}{n_1})+n_2 -m}
.\label{4-25-2}
\end{align}
where 
the inequality (\ref{4-25-1}) follows from (\ref{11-16-2})
and 
the inequality (\ref{4-25-2}) does from 
the following inequality:
\begin{align*}
|\{z||\pi_1(z)- \pi_1(y)|\le t\}| \le 
2^{n_1 \overline{h}(\frac{t}{n_1})+n_2 }.
\end{align*}
Therefore, we obtain (\ref{11-16-6}).
\end{proof}

\section{Proof of (\ref{2-9-2})}\label{2-14-1}
Since $-(1-x)\log (1-x)\le x$,
we have
\begin{align*}
& \overline{h}(P^{\cP}_{ph,min,x|\cD_e,\POS})+ 
\rE^{\cP}_{\cD_e,\POS} 
(
N\eta-m) 
P^{\cP}_{ph,min,x|\cD_e,\POS}\nonumber\\
\le &
-P^{\cP}_{ph,min,x|\cD_e,\POS}
\log( P^{\cP}_{ph,min,x|\cD_e,\POS})
\nonumber\\
&\quad +
P^{\cP}_{ph,min,x|\cD_e,\POS}
+ 
\overline{N}
P^{\cP}_{ph,min,x|\cD_e,\POS}\nonumber\\
= &
P^{\cP}_{ph,min,x|\cD_e,\POS}
(\overline{N}+1-\log P^{\cP}_{ph,min,x|\cD_e,\POS}).
\end{align*}
Thus, using the concavity of the function $x \to - x \log x$,
we obtain (\ref{2-9-2}).

\section{Proof of (\ref{1-31-1-2})}
We will prove (\ref{1-31-1-2}).
First, 
the function 
$q \mapsto f(q):=\sqrt{1-p}\sqrt{q}+\sqrt{p}\sqrt{1-q}$
is monotone increasing for $q \le 1/2$ when $p \le 1/2$.
This is because 
the derivative is calculated as
$f'(q)=
\frac{1}{2}(\sqrt{\frac{1-p}{q}}-\sqrt{\frac{p}{1-q}})\le 0$.
Thus, 
\begin{align*}
&\rE^{\cP}_{M_p,\cD_e,\POS} 
\biggl(
\sqrt{
P^{\cP}_{ph,min,x|M_p,\cD_e,\POS}
}
\sqrt{1-2^{-(\eta N-m)}}\\
&\quad+
\sqrt{1-P^{\cP}_{ph,min,x|M_p,\cD_e,\POS}}
\sqrt{2^{-(\eta N-m)}}
\biggr)^2\\
\le &
\rE^{\cP}_{M_p,\cD_e,\POS} 
\biggl(
\sqrt{
P^{\cP}_{ph,min,x|M_p,\cD_e,\POS}
}
\sqrt{1-2^{-\underline{N}}}\\
& \quad +
\sqrt{1-P^{\cP}_{ph,min,x|M_p,\cD_e,\POS}}
\sqrt{2^{-\underline{N}}}
\biggr)^2\\
\le &
\biggl(
\sqrt{P^{\cP}_{ph,av,x}}
\sqrt{1-2^{-\underline{N}}}\\
&\quad +
\sqrt{1-P^{\cP}_{ph,av,x}}
\sqrt{2^{-\underline{N}}}
\biggr)^2,
\end{align*}
where the last inequality follows from the concavity of 
$\bigl(
\sqrt{P_{ph}}
\sqrt{1-2^{-t}}
+
\sqrt{1-P_{ph}}
\sqrt{2^{-t}}
\bigr)^2$ concerning $P_{ph}$.

\end{document}